\documentclass[pdflatex,sn-standardnature,iicol]{sn-jnl}

\usepackage{rotating}
\usepackage{float}
\usepackage{tabularx}
\usepackage{makecell}
\usepackage{lmodern}
\usepackage{anyfontsize}
\usepackage{amsmath}
\usepackage{cleveref}
\usepackage{booktabs}
\usepackage{comment}
\usepackage{xcolor}
\usepackage{mdwlist}
\usepackage{graphicx}
\usepackage[normalem]{ulem}
\usepackage{siunitx}
\usepackage{array}
\usepackage{soul}

\jyear{2026}%

\theoremstyle{thmstyleone}%
\theoremstyle{thmstyletwo}%

\theoremstyle{thmstylethree}%

\begin{document}

\title{A Performance Model for Hybrid Quantum-Classical Workflows}       
\author[1]{\fnm{Pooja} \sur{Rao}}
\author[2,4]{\fnm{Dimitar} \sur{Trenev}}
\author[1]{\fnm{J\'er\^ome F.} \sur{Gonthier}}
\author[1]{\fnm{Taylor L.} \sur{Patti}}
\author[3,4]{\fnm{Sebastian} \sur{Stern}}
\author[3,4]{\fnm{Tyler} \sur{Takeshita}}
\author[1]{\fnm{Yuri} \sur{Alexeev}}
\author[2,4]{\fnm{Cedric} \sur{Lin}}
\author[1]{\fnm{Sam} \sur{McArdle}}
\author[1]{\fnm{Justin} \sur{Lietz}}
\author[5]{\fnm{Katherine} \sur{Klymko}}
\author[5]{\fnm{Ermal} \sur{Rrapaj}}
\author[6]{\fnm{Norm} \sur{Tubman}}
\author[7]{\fnm{Samuel} \sur{Stein}}
\author[8]{\fnm{Amir} \sur{Shehata}}
\author[8]{\fnm{Travis} \sur{Humble}}
\author[1]{\fnm{Krysta} \sur{Svore}}
\author[2,4]{\fnm{Peter} \sur{Komar}}
\author[1]{\fnm{Elica} \sur{Kyoseva}}
\affil[1]{NVIDIA Corporation, Santa Clara, CA, USA}
\affil[2]{Amazon Braket, Seattle, WA, USA}
\affil[3]{AWS Worldwide Specialist Organization, Seattle, WA, USA}
\affil[4]{AWS Quantum Technologies, Seattle, WA, USA}
\affil[5]{National Energy Research Scientific Computing Center, Lawrence Berkeley National Laboratory, Berkeley, CA, USA}
\affil[6]{NASA Ames Research Center, Moffett Field, CA, USA}
\affil[7]{Pacific Northwest National Laboratory, Richland, WA, USA}
\affil[8]{Oak Ridge National Laboratory, Oak Ridge, TN, USA}

\abstract

Hybrid quantum–classical workflows are expected to underpin practical quantum computing, yet the community lacks a shared framework for reasoning about integration requirements. We introduce a runtime model that decomposes workflow execution into quantum compute, classical compute, and communication costs at two levels of analysis. At the application level, where communication overhead only affects performance, we use a communication-to-computation ratio to quantify its impact on runtime. At the real-time level, when communication overhead determines feasibility, we examine how decoding and feedback latency affect logical operation times and resource requirements. The model evaluates execution with allocated resources, excluding scheduling and resource management. Applied to representative workflows, it shows that the compute-intensive applications, including those requiring HPC resources, are largely insensitive to communication overhead. However, the model identifies when hardware evolution, such as faster quantum processing and greater parallelism, can shift this balance toward communication bottlenecks.

\AddToHook{cmd/printabstract/after}{%
\par\medskip
\leftskip=0pt\rightskip=0pt
\noindent\begin{minipage}{\textwidth}
\normalfont\footnotesize
Pooja Rao and Dimitar Trenev contributed equally as first authors. Peter Komar and Elica Kyoseva are the supervising last authors.\par
Corresponding authors. E-mail: porao@nvidia.com, dtrenev@amazon.com

\par\smallskip
This manuscript has been authored in part by UT-Battelle, LLC,
under contract DE-AC05-00OR22725 with the US Department of
Energy (DOE). The US government retains and the publisher,
by accepting the article for publication, acknowledges that
the US government retains a nonexclusive, paid-up, irrevocable,
worldwide license to publish or reproduce the published form
of this manuscript, or allow others to do so, for US government
purposes. DOE will provide public access to these results of
federally sponsored research in accordance with the DOE Public
Access Plan (\url{https://www.energy.gov/doe-public-access-plan}).
\end{minipage}\par
}

\maketitle
\clearpage
\pagestyle{plain}

\section{Introduction}
\label{sec:intro}

Over the past few decades, advances in classical hardware, software, and algorithms have driven the rapid growth of computational science and enabled leadership-class computing facilities to carry out exascale computations \cite{draeger2024exascaleimpact, allcock2025aurora}. The adoption of heterogeneous computing architectures has been central to this progress, delivering substantial performance gains for applications by leveraging specialized accelerators like graphical processing units (GPUs), which excel in parallelized computing tasks \cite{owens2008, nickolls2008,mittal2015}. More recently, quantum processing units (QPUs) have gained serious traction as upcoming computing accelerators that harness quantum phenomena like superposition, entanglement, and phase-controlled interference to accelerate certain computing tasks as much as exponentially.

Classical computing infrastructure has been instrumental in both the design of ~\cite{sommers2025qpu_simulation} and remote access ~\cite{nguyen2024quantumcloud} to the small number of location-bound quantum devices available today, enabling increasingly ambitious hybrid quantum-classical experiments ~\cite{Alexeev2025Artificial, GPT-QE, minami2025combinatorialgqe, zhao2025qcafqmc, shehata2026bridging, seelam2026QCSC}. As the field advances toward fault-tolerant quantum computation, classical computation plays an important role in quantum error correction, calibration, and end-to-end applications, rendering hybrid quantum–classical workflows increasingly relevant to practical quantum computing.

However, integration of quantum processors into heterogeneous architectures presents a unique challenge compared to classical accelerators because quantum computation is constrained by noisy physical qubits, real-time control, and the layered classical processing required for quantum error correction. Accordingly, we distinguish two levels of analysis for hybrid quantum-classical workflows: (i) the application level, where physical device timing constraints are abstracted away, allowing the developer to reason in terms of logical (but possibly noisy) qubits and gates, and (ii) the real-time level~\cite{stankovic1988}, where the physical characteristics of the quantum device place timing requirements on the real-time execution flow (e.g., classical processing must complete within physical qubit coherence times) and bound the size and scale of executable applications. In the fault-tolerant regime, quantum error correction lifts this constraint at the logical level, decoupling classical processing timescales from physical qubit coherence.

Because the quantum computing stack spans both high-level application workflows and low-level real-time control, the integration architecture connecting classical and quantum resources can differ by orders of magnitude in latency and bandwidth, from remote access over a wide-area network, through co-location within the same facility, to co-designed systems in which quantum and classical resources share a node or are connected through a dedicated low-latency interconnect~\cite{honda2025colocation, Caldwell2025NVQLink}. Recent strategic planning for integrating fault-tolerant quantum computers into HPC facilities similarly distinguishes independent, loosely coupled, and tightly coupled modes of quantum--classical operation~\cite{osti_3022533}, mirroring the latency and bandwidth continuum considered here. Choosing the appropriate integration tier for a given workflow is therefore a critical architectural decision, one that requires understanding whether a workflow can tolerate remote latency, benefits from co-location with classical resources, or requires a co-designed system.

Recent work has begun to map out hybrid quantum-classical integration requirements across system architecture, scheduling, benchmarking, and interconnect~\cite{Caldwell2025NVQLink,mohseni2025position, dobler2025survey, beck2024, Rallis2025, sitdikov2025, zimboras2025eu, khalid2025impacts}, including co-design studies of how device-level parameters and CPU--QPU communication affect runtime on tailored hybrid systems~\cite{mauerer2022}. What is still missing, however, is a shared formalism through which the quantum-algorithms and classical-computing communities can jointly reason about the integration requirements of a given workload. In classical computing, performance models such as LogP~\cite{culler1993logp} and the roofline model~\cite{williams2009roofline} relate workflow characteristics to hardware performance and identify bottlenecks. For multinode quantum computers, Ang et al.~\cite{ang2024arquin} develop a quantum roofline model that relates local computation to internode quantum communication costs. In quantum computing, device-level benchmarks such as CLOPS~\cite{wack2021} and quantum volume~\cite{qvolume2019} characterize quantum hardware performance, while QUOPS~\cite{proctor2026quops} measures computational capability
and throughput across physical and fault-tolerant architectures. Neither provides a workflow-level diagnostic that connects algorithm structure to integration requirements for hybrid quantum-classical systems. Resource-estimation tools estimate quantum-side quantities such as qubit counts, gate counts, logical depth, logical-cycle costs, and quantum runtime under specified architectural assumptions~\cite{AzureQuantumRE,Qualtran2024,harkness2026ftcircuitbench}; the framework introduced here is complementary, embedding such quantum costs in a workflow-level model that also accounts for classical compute and communication structure.

The model applies at both the application level, where it quantifies the performance impact of communication overhead, and the real-time level, where it evaluates whether timing constraints can be met and how the reaction time--the time needed to decode measurement outcomes and apply the resulting correction--shapes fault-tolerant resource requirements.  Applying this model to representative workflows, we show how it informs the assessment of performance across integration tiers, identifying which workloads can tolerate remote latency, which benefit from co-location with classical resources, and how these conclusions may shift as hardware and algorithms evolve. Together, the runtime model and its diagnostics establish a common quantitative language for hybrid quantum-classical integration decisions.
 
The remainder of this paper is organized as follows. Section~\ref{sec:metrics} introduces the runtime model and derives two diagnostics that inform integration decisions: the communication-to-computation ratio (application level) and the real-time feasibility constraint (real-time level). Section~\ref{sec:examples} applies both diagnostics to representative workflows at each level. Finally, Section~\ref{sec:discussion} synthesizes the integration implications and discusses how future developments in hardware and algorithm design may shift integration requirements.

\section{Performance Model}
\label{sec:metrics}

Hybrid quantum--classical algorithms are characterized by alternating phases of quantum and classical computation, coupled through explicit communication events. To enable principled comparisons across workflows and integration architectures, we introduce a runtime model that separates workflow-intrinsic metrics from hardware-dependent parameters, avoiding bias towards any particular qubit modality.

The model addresses a classical host connected to a quantum execution resource, whose endpoints depend on the level of analysis: at the application level, a host ranging from a single workstation to a leadership-class HPC system paired with an application-facing \emph{logical QPU}; at the real-time level, a low-latency control system paired with the physical QPU. The logical QPU may be implemented directly on physical qubits in near-term devices or through fault-tolerant logical qubits in future systems; in the latter case, the real-time classical processing required for error correction is absorbed into this abstraction.

\begin{figure*}[t]                                   
      \centering
      \includegraphics[width=0.85\textwidth]{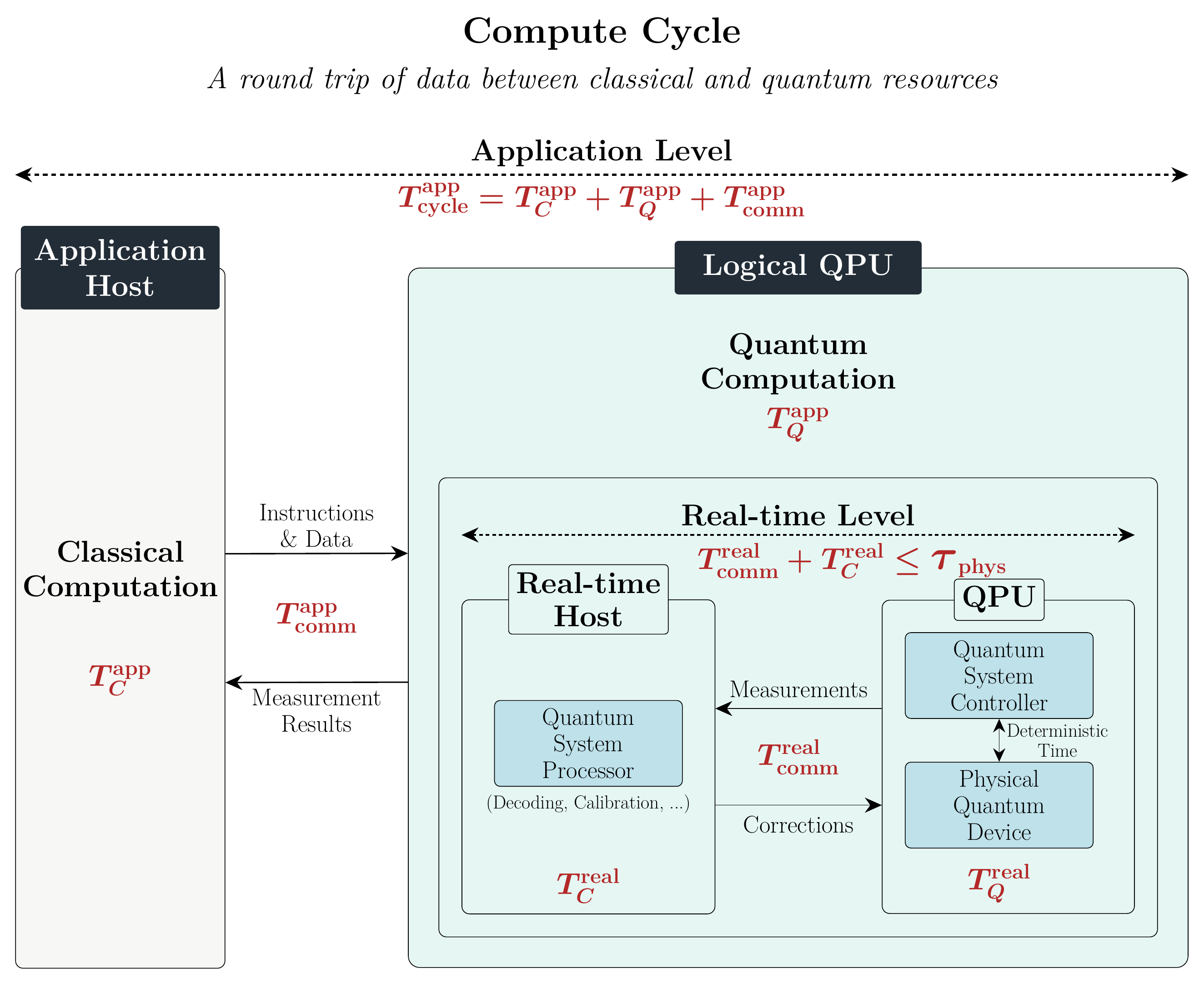}
      \caption{The compute cycle: a complete round trip of information between the classical host and quantum accelerator. The total cycle time $T_{\mathrm{cycle}} = T_C + T_Q + T_{\mathrm{comm}}$ decomposes into classical compute time ($T_C$), quantum compute time ($T_Q$), and communication overhead ($T_{\mathrm{comm}}$). At the real-time level, classical processing and communication must satisfy $T_{C,\mathrm{step}} + T_{\mathrm{comm,step}} \leq \scalebox{1.3}{$\tau$}_{\mathrm{phys}}$ to preserve correctness.}
      \label{fig:compute-cycle}
\end{figure*}

The model is defined per \emph{compute cycle}: the smallest repeated unit of work that includes at least one blocking classical-quantum exchange (e.g., one optimizer iteration, one measurement-and-feedback step). Its precise definition is workflow-specific; Table~\ref{tab:compute-cycle} in Section~\ref{sec:examples} gives the definitions used in this paper. For each compute cycle, we decompose the total time into three contributions:
\begin{equation}
T_{\mathrm{cycle}} = T_C + T_Q + T_{\mathrm{comm}},
\end{equation}
where $T_C$ is the classical compute time, $T_Q$ is the quantum compute time, and $T_{\mathrm{comm}}$ captures blocking quantum-classical communication. Figure~\ref{fig:compute-cycle} illustrates this decomposition and also depicts the real-time level cycle operating within the quantum device, discussed below.

\noindent The separation between $T_Q$, $T_\mathrm{comm}$, and $T_C$ requires clarification, because the boundary depends on what counts as ``operating the quantum device." For instance, a fault-tolerant quantum computation employs classical decoding to correct quantum errors; one could view the decoder as part of operating the quantum device and absorb its cost into $T_Q$. The assignment of device, controller, decoder, service, and network costs to $T_C$, $T_Q$, and $T_{\mathrm{comm}}$ depends on the boundary of the exchange being analyzed. To resolve this ambiguity, we distinguish between two levels of workflow analysis, according to whether the quantum-classical interaction is intrinsically subject to the physical timing constraints of the quantum device.

\textbf{\emph{Application-level}} analysis applies to workflows in which correctness does not require classical processing to complete within any device-imposed timescale. Quantum programs may include adaptive or dynamic circuits, but these adaptations occur only at coarse-grained synchronization points (or pertain to stable logical qubits) and do not require feed-forward within physical coherence times. In the FTQC regime this applies to almost all workflows, as real-time classical feedback needed for logical qubit stability is absorbed into the logical QPU abstraction, increasing the logical operation time rather than the classical computing cost.
 
\textbf{\emph{Real-time level}}  analysis applies to workflows in which classical computation must complete within a timescale set by the physical quantum device to preserve correctness. Here, classical processing consumes measurement outcomes from qubits that remain active and produces control decisions that directly affect ongoing quantum evolution. Quantum error correction and dynamic circuits without full fault tolerance fall under this category.

The distinction is independent of whether a circuit uses classical feed-forward logic (``dynamic'') or not. Dynamic circuits executed on fault-tolerant qubits are application-level workflows; the same circuits executed without fault tolerance require real-time level analysis, since feed-forward must occur within physical qubit lifetimes.

\textbf{Classical compute time ($T_C$)} represents wall-clock processing time for classical work at the level being analyzed. At the application level, it accounts for tasks such as gradient evaluations, optimization updates, tensor contractions, diagonalization, or other post-processing. At the real-time level, the analogous per-step quantity $T_{C,\mathrm{step}}$ accounts for control-path work such as decoding or feed-forward, and affects the feasibility of the workflow instead of the communication-to-computation ratio (defined below).
In order to separate workflow-intrinsic cost from hardware-specific performance, we express
\begin{equation}
T_C = \frac{C_C}{\tau_C},
\end{equation}
where $C_C$ is the workflow-dependent \emph{classical compute cost} (e.g., floating-point operations per cycle) and $\tau_C$ is the classical hardware throughput (e.g., FLOP/s). Parallelization that distributes the same work across more resources increases $\tau_C$; restructuring the algorithm for parallelism (e.g., trading more total operations for shorter wall-clock time) changes $C_C$ and constitutes a different workflow specification. To compare classical computing costs across workflows, it is helpful to anchor them to familiar computational scales: low-cost workloads complete in under a minute on a laptop, medium-cost tasks require hours on a multi-GPU node, and high-cost computations necessitate HPC resources.

\textbf{Quantum compute time ($T_Q$)} depends on both the workload and hardware specifications. To allow hardware-agnostic comparison across modalities, we express
\begin{equation}
T_Q = \frac{C_Q}{\tau_Q},
\end{equation}
where $C_Q$ is the workflow-dependent \emph{quantum compute cost} per cycle and $\tau_Q$ is the hardware-dependent throughput. $C_Q$ does not include device-specific compilation or routing overheads, which are absorbed into $\tau_Q$. A convenient abstraction is
\[
C_Q \propto f(n, d)\cdot s,
\]
where $n$ is the number of logical qubits, $d$ is the logical circuit depth (prior to device-specific compilation), and $s$ is the number of shots required to achieve a target precision. In the simplest model, $f(n, d) = d$ and $C_Q \propto d \cdot s$, in which case $\tau_Q$ is naturally measured in circuit layer operations per second (CLOPS)~\cite{wack2021}. More refined models could account for circuit width or individual gate-counts~\cite{tremba2025circuitdepth}. As with $C_C$, improvements in hardware increase $\tau_Q$ and reduce $T_Q$, but do not change the intrinsic cost $C_Q$. Quantum resource-estimation tools can provide inputs to $C_Q$, $\tau_Q$, or $T_Q$; Appendix~\ref{sec:appendix-metrics} summarizes how such outputs map into the runtime model.

\noindent \textbf{Communication time ($T_{\mathrm{comm}}$)} is evaluated using a standard communication performance model~\cite{hockney1994communication, Lastov2009communication}:
\begin{equation}
T_{\mathrm{comm}} = F \left(L + \frac{V}{B}\right),
\end{equation}
where $F$ is the \emph{communication frequency} (number of blocking quantum-classical exchanges per compute cycle), $V$ is the \emph{data volume per exchange}, $L$ is the round-trip latency between the endpoints relevant to the level of analysis, and $B$ is the available interconnect bandwidth. At the application level, $L$ refers to latency between the classical application host and the logical QPU interface. At the real-time level, we use the boundary shown in Fig.~\ref{fig:compute-cycle}. $T_C$ includes decoding and feedback on the real-time host, $T_{\mathrm{comm}}$ includes communication between the real-time host and controller, and $T_Q$ includes device operations, controller processing, and controller--device signaling. QEC reaction time and any resulting qubit idle time are not added separately. Data transfers that do not gate computation progress are not counted in $F$; their cost is absorbed into $T_Q$ or $T_C$.

We note that $F$ is distinct from the shot count $s$: $s$ determines how many repeated quantum executions are needed for statistical accuracy, while $F$ counts the blocking exchanges that structure the workflow.

\emph{Example:} In VQE, a compute cycle corresponds to one optimizer step, but each group of commuting Hamiltonian terms may require a separate measurement and classical aggregation, leading to $F > 1$ per cycle. With batched submission, $F = 1$. This illustrates that $F$ is a workflow-implementation metric rather than a purely algorithmic one. For workflows where sequential data dependence forces each exchange (e.g., QE-MCMC~\cite{layden2022}), $F$ is irreducible; for workflows with independent shots, it can be minimized through batching.

\textbf{Communication-to-computation ratio ($R_{cc}$).} In classical computing, the communication-to-computation ratio is widely used to predict parallel performance early in the design process~\cite{crovella1992ccr}. A related concept, arithmetic intensity~\cite{williams2009roofline} (FLOPS per byte of data movement), serves as the independent variable in the roofline performance model. That formulation works because all computation is measured in a common unit (FLOPS). In hybrid quantum-classical workflows, computation spans two incommensurable paradigms (classical FLOPS and quantum CLOPS), making a FLOPS/byte-style metric inapplicable. We therefore define $R_{cc}$ as a unitless ratio of wall-clock times:
\begin{equation}
    R_{cc} = \frac{T_{\mathrm{comm}}}{T_Q + T_C} = \frac{F(L + V/B)}{T_Q + T_C}.
\label{eq:rcc}
\end{equation}
When $R_{cc} \gg 1$, the workflow is communication-bound, and reducing the latency $L$ and/or the transfer time $V/B$ can shorten runtime substantially. When $R_{cc} \ll 1$, the workflow is compute-bound and reducing communication overhead has minimal impact on performance.

Two properties are worth noting. First, because $R_{cc}$ is defined in terms of wall-clock times, it is hardware dependent: a tradeoff for generality across compute paradigms. The same workflow can have different $R_{cc}$ values on different hardware, which is precisely what makes it useful for integration decisions. Second, $R_{cc}$ is insensitive to the granularity of the compute cycle definition: if a cycle is redefined to encompass $k$ fine-grained steps, $F$, $C_C$ and $C_Q$ each scale by $k$, and the ratio is unchanged. This holds exactly for homogeneous steps; for heterogeneous workflows, $R_{cc}$ represents a cycle-averaged value.

$R_{cc}$ is primarily an application-level diagnostic. At the real-time level the first question is whether timing constraints can be met at all; once they are, the reaction time continues to act as a performance variable, setting the logical clock speed (Section~\ref{sec:examples}).

\textbf{Real-time feasibility constraint.} At the real-time level, classical processing and communication must complete within a timescale set by the quantum device. Per exchange, the binding condition is
\begin{equation}
    T_{C,\mathrm{step}} + T_{\mathrm{comm,step}} \leq \tau_{\mathrm{phys}},
\label{eq:feasibility}
\end{equation}
where $\tau_{\mathrm{phys}}$ is the relevant physical timescale (e.g., qubit coherence time, logical gate cycle). Whether this condition is satisfied depends on the modality-specific interplay of coherence times, gate speeds, and feed-forward latency. Failure to meet it does not degrade performance; it prevents correct execution entirely.

\textbf{Scope and limitations.} The model evaluates compute-cycle runtime and real-time feasibility for a specified workflow, implementation, and hardware scenario, assuming the required resources are available. It estimates communication sensitivity and provides workload requirements that can inform broader system design studies. It does not select a deployment architecture or produce a procurement recommendation. Job scheduling, admission delays, resource queueing, and co-allocation are outside its scope, as are system capacity, utilization, workload mix, cost, reliability, availability, and operational constraints. The metrics $C_C$ and $C_Q$ are defined for a given algorithmic choice; selecting a fundamentally different algorithm (e.g., a parallel eigensolver in place of a serial one) or a reformulation of the algorithm enabled by new hardware capabilities changes these costs and constitutes a new workflow specification, to be analyzed accordingly.

\section{Illustrative Examples}
\label{sec:examples}

Applying the performance model to concrete workflows requires connecting its metrics to standard quantum algorithm descriptors --- circuit depth, shot count, control flow, feed-forward operations, and classical processing requirements. Workflows that share core primitives or control flow patterns tend to exhibit similar metric profiles, motivating a natural grouping into families (Table~\ref{tab:workflow_class}), from which a representative member can serve as a benchmark for the whole family. We select representatives at both the application and real-time levels that differ along the key axes of the performance model: low vs.\ high classical compute, shot-dominated vs.\ depth-dominated quantum cost, single- vs.\ multi-exchange communication, and low vs.\ high communication-to-computation ratio.

Because the model's metrics ($C_C$, $C_Q$, $F$, $V$, $R_{cc}$) are defined per compute cycle, the choice of compute cycle sets the unit of analysis for each workflow. We take it to be the smallest repeated unit of work that includes at least one classical--quantum exchange (Table~\ref{tab:compute-cycle}). For workflows with independent shots such as Sample-based Quantum Diagonalization (SQD) or the Generative Quantum Eigensolver (GQE), $F$ can be minimized through batching; for workflows with sequential data dependence such as Quantum-Enhanced Markov Chain Monte Carlo (QE-MCMC), $F$ is fixed by algorithmic structure. For the application-level workflows, we estimate runtimes at order-of-magnitude level using reported experimental values where available, identifying whether quantum compute, classical compute, or communication sets the bottleneck. For real-time tasks like QEC, the relevant question is not runtime but whether timing constraints can be met at all; once met, the reaction time remains a performance variable, as the factoring example illustrates. 

Evaluating the communication term additionally requires representative values of latency and bandwidth, which are set by the integration architecture. Integration architectures span a continuum. In remote-access models, the classical host and logical QPU are at different facilities, connected through a wide-area network, and accessed through a cloud service ($L \sim 100$~ms, $B \sim 1$~Gbps). In co-location models, they share a facility and are connected by a direct local-area link ($L \sim 10$~$\mu$s, $B \sim 100$~Gbps)~\cite{seelam2026QCSC}. In co-designed systems, quantum and classical resources reside on the same node or are connected through dedicated low-latency interconnects ($L \sim 100$~ns, $B \sim 500$~Gbps). We refer to these three representative points on the continuum as the remote, co-located, and co-designed integration tiers, and use their order-of-magnitude values to evaluate $R_{cc}$ in the workflows that follow. 

\begin{table}[h]
  \centering
  \caption{Canonical compute cycle definitions used for application-level workflows in this section.}
  \label{tab:compute-cycle}
  \small
  \begin{tabularx}{\columnwidth}{@{}lX@{}}
  \hline
  \textbf{Workflow} & \textbf{Compute cycle} \\
  \hline
  SQD      & One measurement batch \& diagonalization \\
  GQE      & One circuit-batch generation, energy estimation \& weight update \\
  DMRG--QPE & One DMRG sweep schedule, circuit compilation, followed by QPE \\
  QE-MCMC  & One sampling \& accept/reject step \\
  \hline
  \end{tabularx}
\end{table}

\subsection{Application-level workflows}

\begin{figure}[t]
\centering
\includegraphics[width=\columnwidth]{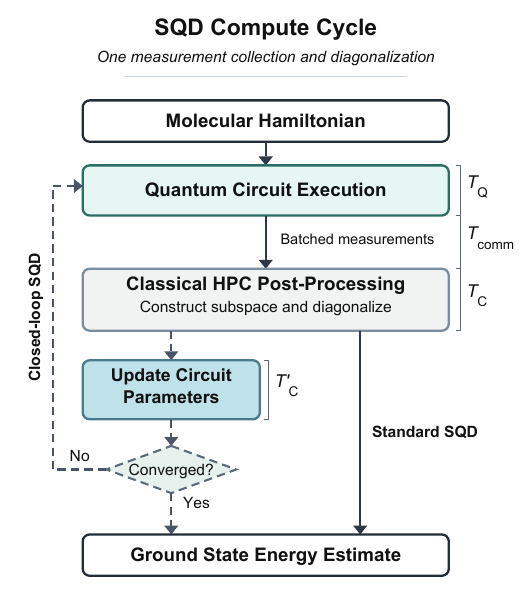}
\caption{Compute cycle for Sample-Based Quantum Diagonalization (SQD). The standard workflow (solid lines) performs a single quantum sampling phase followed by classical subspace construction and diagonalization. The closed-loop variant (dashed lines) iteratively refines the circuit ansatz based on diagonalization results. In both cases, the communication frequency is $F=1$ per iteration, yielding a low communication-to-computation ratio ($R_{cc} \ll 1$).}
\label{fig:sqd-flowchart}
\end{figure}

\paragraph{Subspace Methods: SQD}
SQD~\cite{Kanno2023, robledomoreno2025sqd} leverages shallow quantum circuits together with classical HPC to address larger chemistry problems than previously possible on quantum hardware. The compute cycle is illustrated in Figure~\ref{fig:sqd-flowchart}.

\begin{table}[h]
\centering
\caption{Application-level reference profile for SQD, reported per compute cycle (one measurement \& diagonalization). Reference values from~\cite{robledomoreno2025sqd} (77 qubits, IBM Heron). $\tilde{s}$: subspace dimension after configuration recovery.}
\label{tab:sqd-metrics}
\small
\newcolumntype{L}{>{\raggedright\arraybackslash}X}
\begin{tabularx}{\columnwidth}{@{}lLL@{}}
\toprule
\textbf{Metric} & \textbf{Scaling} & \textbf{Ref.} \\
\midrule
$C_C$ & $O(\tilde{s}^{\,2})$ & $\tilde{s}{\sim}10^{8}$; \newline 90\,min / 64 nodes \\
$C_Q$ & $d{\cdot}s$;\, \newline \mbox{$d{\sim}10^{2}$, $s{\sim}10^{6}$} & $301{\times}2.4$M shots; 45\,min \\
$F$ & $1$ & single batch \\
$V$ & O(s) & $10$--$100$\,MB \\
$R_{cc}$ & $\ll 1$ & ${\sim}10^{-4}$ (remote) \\
\midrule
\multicolumn{2}{@{}l}{Bottleneck} & Classical diag. \\
\bottomrule
\end{tabularx}
\end{table}

The $O(\tilde{s}^{\,2})$ classical cost arises from iteratively solving a generalized eigenvalue problem on a subspace of dimension $\tilde{s}$ constructed from the sampled bitstrings via configuration recovery~\cite{robledomoreno2025sqd}; for recently reported demonstrations, $\tilde{s}$ reaches $10^{8}$, requiring HPC-grade memory. The quantum side is depth-shallow but shot-heavy, with the two-qubit gate count expected to grow only as a low-order polynomial in system size (similar to GQE, discussed below). A single batch of measurements is transmitted per cycle, hence $F = 1$ and the data volume is driven entirely by the shot budget, therefore the quantum sampling and classical diagonalization are two well-separated steps that sequentially run to completion.

We derive $R_{cc}$ explicitly for SQD; subsequent examples follow the same steps. With shallow circuits (${\sim}10^3$ two-qubit gates), depth $d \sim 10^2$ layers, shot budgets $s \sim 10^6$, a single batch exchange per cycle ($F = 1$), and an effective superconducting throughput $\tau_Q \sim 10^5$ layers per second (covering gate execution, measurement, reset, and control overhead), the quantum compute time is
\begin{equation}
    T_Q = \frac{d \times s}{\tau_Q}
        \sim \frac{10^2 \times 10^6}{10^5}
        \sim 10^3\;\text{s}.
\end{equation}
The classical diagonalization has comparable runtime $T_C \sim 10^3$~s at order-of-magnitude level, a precise estimate is unnecessary when $T_C$ does not dominate the sum $T_Q + T_C$. With $F = 1$, the communication-to-computation ratio reduces to
\begin{equation}
    R_{cc}
        = \frac{F (L + V/B)}{T_Q + T_C}
        \sim \frac{L + V/B}{10^3\;\text{s}},
\end{equation}
yielding $R_{cc} \sim 10^{-4}$ at remote-level parameters ($L \sim 10^{-1}$~s, $B \sim 1$~Gbps) with $V \sim 100$~MB per exchange; Table~\ref{tab:sqd-rcc} gives values for the specific application in~\cite{robledomoreno2025sqd}. The workflow is effectively insensitive to communication overhead across any realistic integration tier.

Reconstructing $T_Q$ from the reported experimental parameters in Table~\ref{tab:sqd-metrics} gives $T_Q = (301 \times 2.4 \times 10^6)/(3 \times 10^5) \approx 2{,}400$~s $\approx 40$~min, consistent with the reported $\sim 45$~min. The classical diagonalization contributes $T_C \approx 5{,}400$~s. Table~\ref{tab:sqd-rcc} evaluates $R_{cc}$ across representative integration tiers.

\begin{table}[h]
\centering
\caption{Communication-to-computation ratio of SQD across integration tiers. Parameters are order-of-magnitude estimates provided for illustrative purposes.}
\label{tab:sqd-rcc}
\begin{tabular}{lc}
\toprule
\textbf{Integration tier} & $\boldsymbol{R_{cc}}$ \\
\midrule
Remote ($L = 100$~ms, $B = 1$~Gbps)      & $1.2 \times 10^{-4}$ \\
Co-located ($L = 10$~\si{\micro\second}, $B = 100$~Gbps) & $1.0 \times 10^{-6}$ \\
Co-designed ($L = 100$~ns, $B = 500$~Gbps)     & $2.1 \times 10^{-7}$ \\
\bottomrule
\end{tabular}
\end{table}

$R_{cc} \ll 1$ by at least three orders of magnitude across every tier: the two hours of combined quantum and classical compute dwarf any realistic communication overhead, with even the remote tier (where data transfer dominates over latency) contributing a fractional overhead of $\sim 0.01\%$ to total runtime.

\textbf{Closed-loop variant.} A recent extension~\cite{shirakawa2025closedloop} introduces a classical optimization loop that iteratively updates quantum circuit parameters. Each iteration repeats the full sampling-plus-diagonalization cycle, with subspace dimensions reaching $10^{10}$ and each diagonalization distributed across up to $\sim 38{,}000$ HPC nodes. Despite this substantially higher per-iteration cost, the communication structure remains identical---one quantum sampling phase followed by one classical diagonalization, so $F = 1$ and $R_{cc}$ remains negligible. SQD thus derives no significant benefit from QPU--HPC co-location, even in this more demanding regime.

\begin{figure}[t]
\centering
\includegraphics[width=\columnwidth]{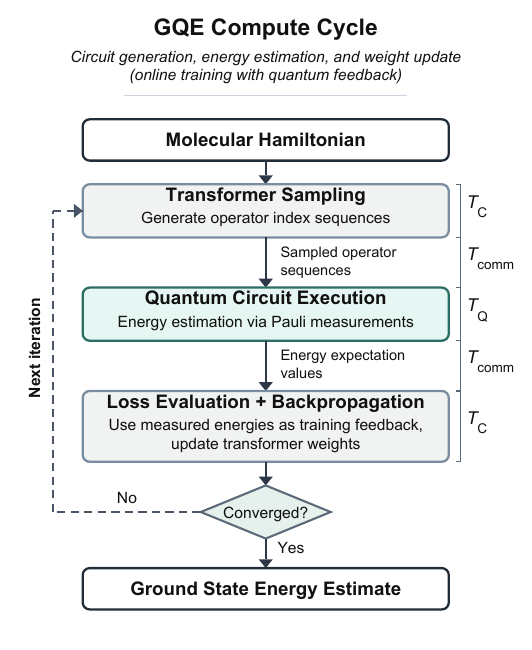}
\caption{Compute cycle for the Generative Quantum Eigensolver (GQE), assuming online training. The transformer samples operator index sequences that define quantum circuits, whose measured energies provide the training feedback. The classical cost is dominated by the backpropagation step; the transformer forward pass (sampling) is less costly, but both contribute to $T_C$. For the GQE configuration considered here, quantum execution dominates the runtime because of the large shot budget required for energy estimation.}
\label{fig:gqe-flowchart}
\end{figure}

\paragraph{AI-based quantum state preparation: GQE}
GQE~\cite{GPT-QE} and its variants use a classical transformer model to generate quantum circuits whose measured energies provide the training signal for the model's weights. We assume online training with quantum data in the loop; each compute cycle comprises generating a batch of circuits sampled from the transformer, estimating their energy on the QPU, and updating the model weights. This loop can start from randomly initialized weights (training from scratch) or continue after pre-training. Pre-training on existing circuit--energy data including classical simulations can reduce the number of training iterations but does not change the per-cycle quantum cost, leaving the workflow compute-bound. The compute cycle is illustrated in Figure~\ref{fig:gqe-flowchart}.

\begin{table}[h]
\centering
\caption{Application-level reference profile for GQE, reported per compute cycle (one optimizer iteration: generate $\to$ estimate $\to$ update). Representative values for a 4-qubit H$_2$ system based on~\cite{GPT-QE}.}
\label{tab:gqe-metrics}
\small
\newcolumntype{L}{>{\raggedright\arraybackslash}X}
\begin{tabularx}{\columnwidth}{@{}lLL@{}}
\toprule
\textbf{Metric} & \textbf{Scaling} & \textbf{Ref.} \\
\midrule
$C_C$ & $O(l\,d_m^2)$; \newline $l$: tokens, \newline $d_m$: model dim. & ${\lesssim}1$~s (GPU) \\
$C_Q$ & $d \cdot s$;\, \newline $d{\sim}10$, $s{\sim}10^6$ & $10{\times}15{\times}8192$ shots; \newline ${\sim}100$~s \\
$F$   & 1 (batched) to \newline circuits $\times$ Pauli terms & 1 (batched) to \newline 150 (per-term) \\
$V$   & ${\sim}4$~kB ($F{=}150$) to \newline ${\sim}600$~kB ($F{=}1$) & \\
$R_{cc}$ & $\ll 1$ & $10^{-3}$--$10^{-1}$ \newline (remote) \\
\midrule
\multicolumn{2}{@{}l}{Bottleneck} & Quantum shot budget \\
\bottomrule
\end{tabularx}
\end{table}

The classical cost per cycle is a transformer forward-plus-backward pass, completing in ${\sim}1$~s on a GPU cluster for circuit sequences of $l \sim 10$--$100$ tokens. On the quantum side, gate depth grows as a low-degree polynomial~\cite{GPT-QE, tyagin2025qaoagpt}, and the shot budget is driven by statistical energy estimation: naive grouping gives $\sim N^2$ scaling in molecular size, while factorization-based measurements reduce this close to linear in $N$~\cite{oumarou2024accelerating, rocca2024reducing}, with prefactors of ${\sim}10^6$ at chemical accuracy~\cite{choi2023fluid, gonthier2022measurements}.

For the 4-qubit H$_2$ system of~\cite{GPT-QE} (depth $d \sim 10$ layers, ${\sim}10^6$ shots per cycle, $\tau_Q \sim 10^5$ CLOPS), we obtain $T_Q \sim 100$~s against $T_C \lesssim 1$~s, making the workflow quantum-dominated. With remote parameters ($L = 100$~ms, $B = 1$~Gbps), $R_{cc}$ ranges from $\sim 10^{-1}$ (per-term measurement communication, $F = 150$) to $\sim 10^{-3}$ (fully batched measurement communication, $F = 1$). Our metrics for GQE are summarized in Table \ref{tab:gqe-metrics}. This places GQE in the same compute-bound regime as SQD: any workflow requiring large shot budgets for Hamiltonian expectation values will accumulate enough quantum compute time to suppress communication overhead. This buffer could shrink as QPU speeds increase or as multiple QPUs distribute the shot budget in parallel.

Our $R_{cc}$ metric allows us to quantify the impact of such potential future developments on the integration requirements for GQE. Here we provide an illustrative example of such an analysis. First, we parametrize the GQE cycle time with two factors that compress $T_Q$ without altering the communication structure: per-device throughput gains and multi-QPU shot parallelism. For GQE, where each shot is an independent circuit execution, distributing the shot budget across $P$ QPUs is straightforward, giving an effective throughput $\tau_Q^{\mathrm{eff}} = P \cdot \tau_Q$. Consider a utility-scale instance targeting a ${\sim}50$-qubit molecule with $d \sim 10^2$ and $s \sim 10^8$ (consistent with chemical-accuracy requirements under factorization-based measurement schemes~\cite{oumarou2024accelerating, rocca2024reducing, gonthier2022measurements}). With $F = 1$ (batched) and $V/B \ll L$ ($V \sim 100$~kB at 1~Gbps gives $V/B \sim 10^{-4}$~s $\ll L = 0.1$~s):
\begin{equation}
    R_{cc} \approx \frac{L}{d \cdot s\,/\,(P \cdot \tau_Q)}.
\end{equation}
Table~\ref{tab:gqe-crossover} evaluates this expression across potential hardware scenarios.

\begin{table}[h]
\centering
\caption{Projected $R_{cc}$ for utility-scale GQE ($d = 10^2$, $s = 10^8$, $F = 1$, $L = 100$~ms) under hardware evolution.}
\label{tab:gqe-crossover}
\begin{tabular*}{\columnwidth}{@{\extracolsep{\fill}}lccc@{}}
\toprule
\textbf{Per-QPU speedup} & $\tau_Q$ & $P$ (QPUs) & $R_{cc}$ \\
\midrule
$1\times$   & $10^5$ & 1    & $10^{-6}$ \\
$10\times$  & $10^6$ & 1    & $10^{-5}$ \\
$1\times$   & $10^5$ & 100  & $10^{-4}$ \\
$10\times$  & $10^6$ & 100  & $10^{-3}$ \\
$100\times$ & $10^7$ & 1000 & $10^{-1}$ \\
\bottomrule
\end{tabular*}
\end{table}

Even under aggressive assumptions, $R_{cc}$ reaches only ${\sim}10^{-1}$ at remote-access latency ($L = 100$~ms), showing that even with future hardware evolutions, GQE is unlikely to need co-location or a co-designed system at the application level. This is because the shot budget acts as a deep computational buffer that keeps the workflow compute-bound across all foreseeable hardware scenarios. Crossover to $R_{cc} \gtrsim 1$ would require shot-budget reductions of several additional orders of magnitude (e.g., via amplitude estimation \cite{knill2007optimal}) or a qualitative change in communication structure forcing $F \sim s$.

\paragraph{Fault-tolerant ground-state energy estimation:
  DMRG-assisted QPE}
  \label{sec:dmrg-qpe}
We consider qubitization-based quantum phase estimation (QPE) for the ground-state energy of $^{143}$Ce to three-decimal-digit resolution, in a nuclear shell-model space of $N_{\mathrm{sys}}=76$ single-particle orbitals~\cite{gibbs:2026, yoshida:2026}. One compute cycle comprises classical DMRG and circuit compilation, a batch of approximately four QPE executions, and classical phase readout. We assume no inter-shot classical feedback is required, so the batch is submitted in one blocking application-level exchange ($F=1$). Classical DMRG and compilation are performed once; the quantum state-preparation circuit is executed before each QPE shot.

\begin{table}[h]
\centering
\caption{Application-level profile for DMRG-assisted QPE,
per batched compute cycle. $N_T$ excludes state preparation.
$^\dagger$Under the illustrative $T_C$ estimate, which assumes
compilation and classical readout are subdominant.}
\label{tab:qpe-metrics}
\small
\newcolumntype{L}{>{\raggedright\arraybackslash}X}
\begin{tabularx}{\linewidth}{@{}lLL@{}}
\toprule
\textbf{Metric} & \textbf{Scaling} & \textbf{Ref.} \\
\midrule
$C_C$
& $O(N_{\mathrm{iter}}N_{\mathrm{sys}}\chi^3)$ (DMRG);
  \newline
  $O(N_{\mathrm{sys}}\chi_{\mathrm{comp}}^2)$ (compilation model)
& $\chi=10^3$;
  \newline $\chi_{\mathrm{comp}}=256$;
  \newline $N_{\mathrm{iter}}=50$;
  \newline H100 (80~GB) for DMRG \\
$C_Q$
& $s(N_T+N_{T,\mathrm{prep}})$;
  \newline $N_T\sim10^{10}$, $s\sim4$
& $\sim4\times10^{10}$ T gates \\
$F$
& $1$
& Batched executions \\
$V$
& Block-encoding coefficients and state-preparation instructions
& $\sim100$~MB (assumed) \\
$R_{cc}$
& $\displaystyle\frac{L+V/B}{T_Q+T_C}$
& $\lesssim2.25\times10^{-6}$ (remote) \\
\midrule
\multicolumn{2}{@{}l}{Bottleneck}
& Quantum execution$^\dagger$ \\
\bottomrule
\end{tabularx}
\end{table}

Ref.~\cite{gibbs:2026} uses an H100 GPU (80~GB) for DMRG at bond
dimension $\chi=1000$. In the implementation of Ref.~\cite{gibbs:2026}, the circuit-compilation code did not exploit GPU acceleration; a reduced MPS with
$\chi_{\mathrm{comp}}=256$, obtained by reverse-sweep
DMRG, was therefore used to improve tractability. The resulting circuits contain $N_{T,\mathrm{prep}}=O(10^4)$ T gates~\cite[Fig.~9]{gibbs:2026}. Assuming $N_{\mathrm{iter}}=50$ iterations, the DMRG scaling term in Table~\ref{tab:qpe-metrics} gives $50\times76\times(10^3)^3=3.8\times10^{12}$. For compilation, we retain a rough quadratic-in-bond-dimension cost model: $N_{\mathrm{sys}}\chi^2=7.6\times10^7$ at $\chi=1000$,
or $N_{\mathrm{sys}}\chi_{\mathrm{comp}}^2\approx5\times10^6$ for the reduced input. These are scaling estimates, not measured
operation counts; the total variational compilation cost also
depends on its optimization procedure. Using the leading DMRG scaling as an order-of-magnitude operation count and assuming an effective H100 throughput of $10^{12}$ FLOP/s gives $T_{\mathrm{DMRG}}\sim4$~s. We adopt $T_C\sim4$~s as an illustrative DMRG-dominated estimate, assuming compilation and classical readout contribute subdominantly.

In addition to the $N_{\mathrm{sys}}=76$ system qubits, we use
$N_{\mathrm{QFT}}=10$ phase qubits for the three-digit case, an LCU index register of $N_{\mathrm{LCU}}=\lceil\log_2N_p\rceil\approx17$ qubits, where $N_p$ counts Pauli terms, and additional workspace ancillas for the qubitized walk operator, whose number depends on the number of Pauli terms and the precision used to represent their coefficients~\cite{yoshida:2026}. The estimate uses $N_T\sim10^{10}$ logical T gates per QPE execution, excluding state preparation. Ref.~\cite{gibbs:2026} estimates that these state-preparation circuits reach an overlap magnitude of at least $0.5$ with the exact ground state, so each QPE execution succeeds with probability ${\approx}\,0.25$ and we adopt $s\sim4$. The quantum workload is therefore dominated by the cost of each QPE execution, rather than by a large shot count. We express $C_Q$ in logical T gates and $\tau_Q$ as the effective logical T-gate consumption rate. Assuming a serial execution rate of $\tau_Q=10^5$ T gates/s ($10\,\mu$s per T gate), as an illustrative hardware scenario considered in Ref.~\cite[Sec.~6]{LitinskiGameOfSurfaceCodes}, we estimate
\begin{equation}
    \begin{aligned}
        C_Q &= s\bigl(N_T+N_{T,\mathrm{prep}}\bigr)
             \sim4\times10^{10}\ \text{T gates},\\
        T_Q &= \frac{C_Q}{\tau_Q}
             \sim\frac{4\times10^{10}}{10^5}
             =4\times10^5\ \mathrm{s}.
    \end{aligned}
\end{equation}

We assume that the circuit sent to the QPU is specified compactly by the block-encoding coefficients and the state-preparation circuit rather than gate by gate. Taking $V=10$~MB as an illustrative baseline, we increase the assumed payload tenfold to $V=100$~MB for a more conservative estimate of communication overhead. For fixed latency, bandwidth, and compute times, this larger payload increases $T_{\mathrm{comm}}$ and hence the upper bound on $R_{cc}$. The data returned is the measured phase register, a few bytes per shot. At remote-tier parameters $L=0.1$~s and $B=1$~Gbps,
\[
    T_{\mathrm{comm}}
    =0.1~\mathrm{s}
     +\frac{8\times10^8~\mathrm{bits}}
            {10^9~\mathrm{bits/s}}
    =0.9~\mathrm{s}.
\]
Using the resource estimates above and the assumed serial
execution rate and communication parameters, $T_C\geq0$
gives the following approximate upper bound from
Eq.~\eqref{eq:rcc}:
\begin{equation}
    \begin{aligned}
        R_{cc}
        &=\frac{T_{\mathrm{comm}}}{T_Q+T_C}\\
        &\lesssim\frac{0.9~\mathrm{s}}{4\times10^5~\mathrm{s}}
         =2.25\times10^{-6}.
    \end{aligned}
    \label{eq:dmrg-qpe-rcc}
\end{equation}

High-latency (remote) communication contributes negligibly to runtime for these inputs. Under the illustrative $T_C$ estimate, quantum execution dominates; its rate depends on local control, gate scheduling, and magic-state supply~\cite{LitinskiGameOfSurfaceCodes}.

Returning to the current landscape, to illustrate the opposite extreme, we turn to a workflow built around single-shot quantum-classical exchanges.

\paragraph{Iterative Quantum Markov Sampling: QE-MCMC}
The Quantum-Enhanced Markov Chain Monte Carlo (QE-MCMC)~\cite{layden2022} algorithm accelerates sampling from complex probability distributions, such as the Boltzmann distribution of classical Ising models. A compute cycle is one complete MCMC step: the quantum processor prepares and measures a proposal state, the result is sent to the classical host to be accepted or rejected, and the decision sets up the next iteration.

\begin{table}[h]
\centering
\caption{Application-level reference profile for QE-MCMC, reported per compute cycle (one sampling \& accept/reject step). Experimental values from~\cite{layden2022} (10 qubits, IBM superconducting).}
\label{tab:qemcmc-metrics}
\small
\newcolumntype{L}{>{\raggedright\arraybackslash}X}
\begin{tabularx}{\columnwidth}{@{}lLL@{}}
\toprule
\textbf{Metric} & \textbf{Scaling} & \textbf{Ref.} \\
\midrule
$C_C$ & $O(n)$ (sparse) to \newline $O(n^2)$ (fully connected) & $T_C \sim 10^{-8}$~s \\
$C_Q$ & $d \cdot 1$;\, single shot & 2--48 two-qubit gate \newline layers; $T_Q \sim 10^{-4}$~s \\
$F$   & 1 (irreducible) & \\
$V$   & $n$ bits + 1 bit & ${\sim}10$~B \\
$R_{cc}$ & $\gg 1$ (remote) & ${\sim}10^3$ (remote); \newline ${\sim}10^{-1}$ (co-located); ${\sim}10^{-3}$ (co-designed) \\
\midrule
\multicolumn{2}{@{}l}{Bottleneck} & Round-trip latency \\
\bottomrule
\end{tabularx}
\end{table}

Both the classical and quantum costs per step are minimal: the classical host evaluates an energy difference and a Metropolis acceptance criterion in a handful of arithmetic operations, while the quantum processor executes a shallow mixing circuit and returns a single measured bitstring. With $F = 1$ and negligible $T_C, T_Q$, the single round-trip dominates the cycle under remote access: applying Eq.~\eqref{eq:rcc} with the representative latencies from Table~\ref{tab:sqd-rcc}, remote access gives $R_{cc} \sim 10^3$, while co-location ($\sim 10^{-1}$) and co-designed systems ($\sim 10^{-3}$) bring the workflow into the compute-bound regime. This places QE-MCMC at the opposite end of the communication-to-computation spectrum from SQD.

The severity of this bottleneck is reflected in practice. The authors of~\cite{layden2022} did not implement the algorithm as an online iterative loop, reporting that classical infrastructure round-trip latency would have made online execution impractically slow, which the $R_{cc} \gg 1$ diagnostic identifies. Instead, quantum data was collected offline in batch and MCMC chains were extracted by subsampling cached transitions, a strategy that eliminates the sequential exchange but does not scale well to large systems. The authors identify dynamic circuit capabilities, where classical feed-forward logic executes within the quantum control system rather than round-tripping to an external host, as the enabling technology for practical online execution. Subsequent implementations that batch many shots per iteration and increase per-cycle classical processing~\cite{marshall2026mcmcoptim} shift the workflow from latency-bound toward compute-bound, reducing $R_{cc}$ as per-cycle compute absorbs the communication overhead. Table \ref{tab:qemcmc-metrics} summarizes our metrics for QE-MCMC.

\medskip
The four application-level workflows above span a wide range of communication-to-computation ratio, from the deeply compute-bound SQD and DMRG+QPE, through GQE, whose shot budget absorbs communication overhead, to the communication-dominated QE-MCMC. For these application-level workflows, $R_{cc}$ quantifies the runtime overhead due to communication. Figure~\ref{fig:rcc-network} in the appendix shows how this overhead depends on the workflow and integration tier: under the stated parameters, SQD and GQE remain compute-bound, whereas communication dominates QE-MCMC under remote access. At the real-time level, correct quantum execution requires classical processing and communication to meet device-imposed timing constraints. At the real-time level, the question first shifts to feasibility: can classical processing and communication complete within a timescale set by the physical device?

Dynamic-circuit workflows, such as SQSP-MaF in Appendix~\ref{sec:appendix-sqsp}, illustrate the boundary between the two regimes. The relevant level of analysis is determined not by the presence of mid-circuit measurement and feed-forward alone, but by where that feed-forward is implemented. When executed directly on physical qubits, measurement-dependent corrections must occur within the physical coherence window, making the binding constraint real-time feasibility. When implemented on fault-tolerant logical qubits, this constraint is absorbed into the logical QPU abstraction, and the workflow reverts to application-level analysis.

\subsection{Real-time level workflows}

Quantum error correction (QEC) is crucial to achieving fault-tolerant quantum computation (FTQC), but brings significant challenges in quantum control systems and classical computation. A common motif across leading FTQC candidates is syndrome extraction measurements yielding bits of information that can be used to diagnose and correct errors; processing this syndrome data to infer the underlying errors is known as syndrome decoding (or simply decoding). We define one compute cycle as one stabilizer measurement cycle followed by applying corrective operations if needed.

\begin{table}[h]
\centering
\caption{Real-time level reference profile for QEC, reported per compute cycle (one stabilizer measurement followed by correction if needed), except that $F$ is expressed as a streaming rate because stabilizer cycles run continuously rather than as discrete batches.}
\label{tab:qec-metrics}
\small
\newcolumntype{L}{>{\raggedright\arraybackslash}X}
\begin{tabularx}{\columnwidth}{@{}lL@{}}
\toprule
\textbf{Metric} & \textbf{Value} \\
\midrule
$C_C$    & Decoder cost trades accuracy / throughput / latency; grows with code distance, complexity, and logical operations. AI decoders now meet real-time throughput for superconducting qubits (demonstrated to distance 11 for the surface code)~\cite{senior2026scalablerealtimeneuraldecoder} \\
$C_Q$    & Shallow syndrome-extraction circuits; few samples; for a $d=5$ rotated surface code, ${\sim}80$ two-qubit gates and ${\sim}24$ stabilizer measurements~\cite{google2023suppressing} \\
$F$      & Continuous streaming: ${\sim}1/$\si{\micro\second} (superconducting, spin, photonic); ${\sim}1$/ms (atom, ion) \\
$V$      & ${\sim}1$ bit per qubit per round; aggregate can reach MB at MHz, PB per program for 1M-qubit surface code \\
$R_{cc}$ & Replaced by the feasibility condition (Eq.~\ref{eq:feasibility}) \\
\midrule
\multicolumn{1}{@{}l}{Bottleneck} & Decoder reaction time \\
\bottomrule
\end{tabularx}
\end{table}

Decoders are characterized by three coupled metrics: \textit{accuracy} (logical error rate), \textit{throughput} (syndrome processing rate), and \textit{decoder latency} (time to compute a correction from syndrome data). Throughput must match or exceed the syndrome generation rate to avoid a computational backlog~\cite{TerhalQECMemories}; parallel decoding architectures now achieve this~\cite{AlibabaParallelTime, RiverlaneParallelTime, khalid2025impacts}. The remaining binding constraint is decoder latency which, unlike the communication latency in HPC contexts, refers primarily to classical computation time rather than data transfer. Combined with control-system communication delays, decoder latency determines the total \textit{reaction time}: the end-to-end delay from measurement to applied correction, which directly sets the logical clock speed~\cite{FowlerGidneyLattice, khalid2025impacts}. Quantum algorithms can be compiled to be reaction-time limited~\cite{fowler2012timeoptimal, GidneyFactor}, meaning the logical clock speed is determined only by the speed of qubit measurement (in the $X$ or $Z$ basis), and how fast the decoder and control stack can react. A full discussion of decoder latency is deferred to Appendix~\ref{sec:appendix-decoding}.

In our framework, the binding QEC constraint at the real-time level is that the reaction time not exceed the logical gate cycle: 
\begin{equation}
    T_{\mathrm{comm}} + T_{\mathrm{decode}} \lesssim T_{\mathrm{logical\_gate}},
\end{equation}
failing which the QPU idles and the effective logical clock drops to the reaction time of the decoder plus control system. This timescale varies widely: microseconds on superconducting devices, milliseconds on trapped-ion and neutral-atom devices~\cite{chen2026realtimeqec}. The compute-volume scaling is severe in the aggregate: for 1M qubits in a surface code, roughly half generate a measurement every microsecond, producing megabytes of syndrome data per millisecond and petabytes over a single quantum program. Decoding and controlling this stream in real time is HPC-scale in aggregate: DOE planning estimates as much as 50 petaflops, at 1 Tb/s bandwidth and 20 $\mu$s latency, for the embedded classical layer~\cite{Caldwell2025NVQLink, osti_3022533}. This load is spatially parallel, however, since syndrome decoding factorizes across logical-qubit patches; it is met by many small, local decoders (FPGAs or ASICs) placed in close proximity to the device rather than a monolithic supercomputer, requiring only bounded coordination between neighboring patches during lattice surgery~\cite{lin2025spatial}. Our metrics for QEC are summarized in Table \ref{tab:qec-metrics}.

\begin{table*}[h!]
    \centering
    \begin{tabular}{c|cc|cc}
         Compilation emphasis & \multicolumn{2}{c|}{Space}
         & \multicolumn{2}{c}{Spacetime} \\
         Reaction time & $10$\si{\micro\second} & $20$\si{\micro\second} & $10$\si{\micro\second} & $20$\si{\micro\second} \\ \hline
        Physical Qubits $(10^6)$ & $18.5$  & $15.3$ & $32.7$  & $26.8$ \\
        Toffoli count $(10^9)$   & $2.7$ & $3.0$  & $3.1$  & $3.2$ \\
        Runtime (Hours)         & $8.1$  & $16.1$  & $3.2$ & $5.3$ \\
        Spacetime volume (Megaqubit-hours) & $150$ & $246$ & $105$ & $142$
    \end{tabular}
    \caption{Resource estimates for using Shor's algorithm~\cite{GidneyFactor} to factor a 2048-bit RSA integer. Resource estimates are obtained using the resource estimate calculator from Ref.~\cite{GidneyFactor}, assuming surface code error correction with physical gate error rate $10^{-3}$, and surface code cycle time $1$\si{\micro\second}.}
    \label{tab:FactoringEstimates}
\end{table*}

Even when the feasibility condition is met, reaction time remains a continuous performance variable: faster decoding means a faster logical clock and fewer resources per algorithm, though is bound by the syndrome data generation rate. The impact depends on circuit structure: Clifford-bottlenecked subroutines (e.g., lookup tables) are largely insensitive to reaction time, while non-Clifford-bottlenecked ones (e.g., adders, rotation synthesis) are directly limited by it~\cite{LitinskiGameOfSurfaceCodes,gidney2019flexible,mcardle2025fast}. We illustrate this with two examples. \\

\noindent{\textit{\textbf{Pauli-based computation (PBC)}} } \\
PBC is a compilation strategy often referring to the process of commuting all Clifford operators out of a circuit, resulting in a circuit comprising solely of non-Clifford Pauli-product rotations \cite{FowlerGidneyLattice}. Architectures that support Pauli-product measurements (PPM), such as lattice surgery over surface code or the gross codes method of computation, are well suited to implement PBC. Operating under the assumption of $T$-states as the magic state resource, resultant circuits comprise layers of PPMs. Each layer is defined as a set of commuting PPM operators, and layers are separated by anti-commuting operators. Each PPM operator requires a $T$-state, and the joint PPM and $T$ measurement determines whether the corrective operation needed. The decoder must resolve this before the next layer can proceed~\cite{fowler2012timeoptimal}. Consider a generic Clifford + $T$ circuit implemented via PBC in the surface code, as described in Ref.~\cite{LitinskiGameOfSurfaceCodes}. The circuit can be modeled as a sequence of $\pi/8$ Pauli-product measurements (PPMs), each acting on all register qubits and a fresh $T$ magic state. Implementing each PPM via lattice surgery takes $d \tau_{se}$ time, where $d$ is the code distance, and $\tau_{se}$ is the time taken for one round of syndrome extraction. We also assume the reaction time of each PPM is $\tau_r$, to determine whether a corrective Clifford operation is required. In general, we must know the Clifford frame before starting the next operator that anticommutes with the correction. Therefore, the effective time per logical operation is $\tau_r + d \tau_{se}$, assuming a sufficient rate of magic state production. We can amortize the overhead if PPMs occur in commuting groups. In this case, we only incur a full reaction time between groups, rather than within them. The effective time per logical operation is reduced to $\frac{\tau_r}{c} + d \tau_{se}$ where $c$ is the size of each commuting group. For example, Ref.~\cite{kim2022Battery} observes that for a family of quantum chemistry calculations, $c \approx 15$. \\

\noindent{\textit{\textbf{Factoring}}}  \\
We illustrate the resource impact of reaction time on a concrete fault-tolerant algorithm: factoring a 2048-bit RSA integer via Shor's algorithm, using the compilation in Ref.~\cite{GidneyFactor}. This compilation balances Clifford-bottlenecked table lookups (insensitive to reaction time) and reaction-limited adders, with the relative cost controlled by lookup size and adder runway separation.

In Table~\ref{tab:FactoringEstimates}, we show resource estimates for compilations of Shor's algorithm for factoring a 2048-bit integer~\cite{GidneyFactor}. All variations use the reaction-limited linear-depth adder circuit~\cite{gidney2019flexible}. This choice already prioritizes runtime, and incurs a higher qubit overhead than a truly space-minimal compilation. This is responsible for the artifact where increasing reaction time appears to reduce physical qubit count, because of a reduced demand for magic state distillation. In a truly space minimizing computation, this would not be the case. Conditioned on reaction-limited addition, we observe that increased reaction time contributes to a large increase in runtime, and in overall spacetime volume. More broadly, the resource impact of reaction time cascades: a faster decoder shortens runtime, which reduces the number of idle cycles during which errors accumulate, which in turn allows a lower code distance to meet the same logical error budget, ultimately reducing the physical qubit count.

\section{Discussion and Outlook}
\label{sec:discussion}

The performance model introduced in this paper provides a shared formalism for assessing hybrid quantum-classical integration requirements at the application and real-time levels. At the real-time level, low latency is first and foremost a threshold requirement: classical processing and communication must complete within the device timescale, and failure prevents correct execution entirely. At the application level, communication latency affects runtime without directly affecting correctness. We distinguish between these levels to clarify how latency, bandwidth, and throughput shape performance under different computational and timing constraints.

Across the application-level workflows analyzed in this work, none are simultaneously compute-intensive and highly communication-bound under the assumed hardware and communication parameters. This is almost by construction, as heavy computation absorbs communication overhead, unless specific hardware or implementation choices (e.g., hardware accelerators, multi-QPU parallelism) sufficiently compress both the classical and quantum computation time. Application-level workflows therefore tend to fall into two categories, as those examined here illustrate: those with heavy per-cycle classical or quantum compute (e.g., SQD, GQE, DMRG-QPE) exhibit low $R_{cc}$ and tolerate remote access, while communication-dominated workloads (e.g., QE-MCMC) require low-latency integration but not HPC resources. For the compute-bound application-level examples considered here, reducing communication overhead offers limited runtime benefit. Barring much faster logical clocks or much larger algorithmic outputs, this assessment is expected to hold under FTQC; DOE planning for fault-tolerant quantum--HPC integration~\cite{osti_3022533} projects that useful FTQC subroutines will require many ($>10^4$) logical clock cycles at slow logical clock speeds ($<100$~kHz) to yield only $10$--$10^3$ bits of output, a relatively long quantum compute time for a small amount of data. At the real-time level, tasks such as QEC require classical processing and communication to meet the timing constraints of the quantum device.

Beyond changes to the algorithmic structure of the workflow, the underlying hardware shifts $R_{cc}$ for the same workflow. Compressing the compute time by adding resources, such as distributing shots across multiple QPUs or parallelizing classical post-processing across more nodes, raises $R_{cc}$ and can move currently compute-bound workflows toward the communication-bound regime. This shot-level parallelism is distinct from coherent distributed quantum computation, in which multiple QPUs jointly execute a single quantum computation through inter-node entanglement or quantum interconnects~\cite{Caleffi2024,Beals2013,Main2025}; such systems change the workflow specification itself and should be modeled separately.

Turning to the real-time level, reaction time matters beyond feasibility alone. The feasibility condition (Eq.~\ref{eq:feasibility}) remains the binding constraint; once it is met, however, reaction time becomes a continuous performance variable whose impact cascades through the full resource stack. Faster decoding shortens runtime, reduces idle-error accumulation, and permits lower code distances --- concretely, doubling the reaction time increases the factoring runtime by 60--100\% depending on the compilation (Table~\ref{tab:FactoringEstimates}). This sensitivity is set by the compilation rather than the algorithm alone: Clifford-bottlenecked subroutines are largely insensitive, while amortizing reaction time over commuting operations, as in Pauli-based computation, reduces the effective cost per logical operation. Because reaction time determines the throughput $\tau_Q$ of the logical QPU, it directly impacts the cost of every application-level workflow executed on logical qubits.

As quantum systems mature, several developments are likely to reshape the compute and communication landscape. Fault-tolerant quantum processors will trade throughput for higher fidelity (QEC overhead reduces $\tau_Q$, but workflows may require fewer shots or exchanges per cycle)~\cite{khalid2025impacts}; the net effect on $R_{cc}$ will be workflow-dependent. At the same time, new algorithms may be designed specifically to exploit low-latency quantum-classical interaction at the application level, much as GPU-aware algorithms were developed to exploit the capabilities of classical accelerators~\cite{nickolls2008}. Finally, the emergence of quantum memory could enable stateful execution models with capabilities inaccessible today~\cite{liu2023quantummemory}, such as state checkpointing, conditional execution on preserved quantum states, and preservation of expensive to prepare quantum states across algorithmic phases. Such capabilities could fundamentally alter the communication structure of workflows and shift the compute-versus-communication balance. The model introduced here renders such assessments more concrete. For example, the hardware-evolution analysis of GQE (Section~\ref{sec:examples}, Table~\ref{tab:gqe-crossover}) identifies the specific conditions (shot-budget reductions of several orders of magnitude or a structural change in communication pattern) under which integration requirements would shift, providing workflow-specific inputs to broader system design studies.

On the classical side, computing requirements tend to grow rapidly with the size of the quantum system (state preparation, Hamiltonian construction, measurement post-processing, error mitigation) even when the quantum algorithm itself is designed to reduce asymptotic complexity. As brute-force methods become infeasible at scale, AI and machine learning are emerging as effective tools for tackling otherwise intractable classical sub-problems, a trend demonstrated by workflows like GQE, which utilize generative models as core elements~\cite{Alexeev2025Artificial}. However, these AI-driven approaches are themselves computationally demanding, requiring HPC-scale compute. The model's separation of workflow-intrinsic costs ($C_C$) from hardware throughput ($\tau_C$) ensures the framework remains applicable as these classical methods evolve.

\backmatter
\section{Contributions}
E.K., P.K., P.R., D.T., T.L.P., Y.A., C.L., and J.F.G. conceived the project. D.T., P.R., J.F.G., C.L., T.L.P., Y.A., J.L., and S.M. carried out the modeling and analysis. P.R., D.T., Y.A., and J.L. prepared the figures. P.R., D.T., T.L.P., J.F.G., T.T., Y.A., K.K., E.R., N.T., Se.S., Sa.S., A.S., J.L., and S.M. wrote the manuscript with input from E.K., P.K., K.S., and T.H. Peter Komar and Elica Kyoseva supervised the work as last authors. Pooja Rao and Dimitar Trenev contributed equally as first authors.

The authors would also like to thank Marwa Farag, Monica Van Dieren, Stefano Mensa, Ikko Hamamura, Damian S. Steiger and Martin Schuetz for feedback and discussions.

\backmatter
\section{Competing interests}
The authors declare no competing interests.

\section{Data Availability}
No new experimental data were collected in this study. The numerical estimates are based on the analytical model, assumptions, and parameter values described in the main text and appendix. Sources of values taken from the literature are cited at the relevant locations.

\bibliography{bibliography}

\backmatter
 \twocolumn
 \begin{appendices}
\section*{Appendix}
\makeatletter
\renewcommand{\@seccntformat}[1]{%
  \csname the#1\endcsname\quad
}
\renewcommand{\sectionfont}{%
  \normalfont\fontsize{12bp}{14bp}\selectfont
  \bfseries\raggedright\boldmath
}
\makeatother
\renewcommand{\theHtable}{B\arabic{table}}
\renewcommand{\theHfigure}{appendix.\thesection.\arabic{figure}}
\renewcommand{\theHequation}{appendix.\thesection.\arabic{equation}}

\section{Model Parametrization}
\label{sec:appendix-metrics}

This section consolidates the definitions, scoring guidelines, and hardware parameters that support the performance model introduced in Section~\ref{sec:metrics}. Table~\ref{tab:glossary} lists all symbols used in the paper. Table~\ref{tab:metric-scale} maps each metric to a representative numerical range on a 1--5 scale, and Table~\ref{tab:metric-profiles} summarizes the resulting profiles for the workflows analyzed in Section~\ref{sec:examples}.

\begin{table*}[!tp]
\tablebodyfont
\caption{Summary of metrics and parameters introduced in this work.}
\label{tab:glossary}
\centering
\renewcommand{\arraystretch}{1.4}
\begin{tabular}{@{} l l p{8.5cm} @{}}
\toprule
\textbf{Symbol} & \textbf{Name} & \textbf{Definition} \\ \midrule
\multicolumn{3}{l}{\textit{Structural}} \\
 & Compute cycle & Recurring unit requiring quantum--classical synchronization \\
$n$ & Qubits & Logical/physical qubits required by the workflow \\
$d$ & Circuit depth & Logical/physical depth, prior to device-specific compilation \\
$s$ & Shot count & Circuit repetitions per compute cycle \\
\midrule
\multicolumn{3}{l}{\textit{Workflow-intrinsic metrics}} \\
$C_C$ & Classical compute cost & Algorithmic cost per cycle on the classical processor \\
$C_Q$ & Quantum compute cost & Algorithmic cost per cycle on the QPU; $C_Q \propto d \cdot s$ \\
$F$ & Communication frequency & Blocking quantum--classical exchanges per cycle \\
$V$ & Data volume per exchange & Classical data transmitted per exchange \\
\midrule
\multicolumn{3}{l}{\textit{Hardware-dependent parameters}} \\
$\tau_C$ & Classical throughput & Classical processing rate (e.g., FLOPS) \\
$\tau_Q$ & Quantum throughput & Quantum execution rate (e.g., CLOPS or QUOPS) \\
$L$ & Round-trip latency & Latency between level-specific endpoints: application host $\leftrightarrow$ logical QPU interface, or real-time host $\leftrightarrow$ quantum system controller + physical quantum device \\
$B$ & Bandwidth & Sustained interconnect data transfer capacity \\
$F_{\mathrm{gate}}$ & Gate fidelity & Average error per gate \\
\midrule
\multicolumn{3}{l}{\textit{Derived quantities}} \\
$T_C$ & Classical compute time & $C_C / \tau_C$ \\
$T_Q$ & Quantum compute time & $C_Q / \tau_Q$ \\
$T_{\mathrm{comm}}$ & Communication time & $F(L + V/B)$ \\
$T_{\mathrm{cycle}}$ & Total cycle time & $T_C + T_Q + T_{\mathrm{comm}}$ \\
$R_{cc}$ & Communication-to-computation ratio & $T_{\mathrm{comm}}/(T_Q + T_C)$; $\gg 1$: communication-bound, $\ll 1$: compute-bound \\
\bottomrule
\end{tabular}
\end{table*}

Table~\ref{tab:resource-estimator-map} summarizes how outputs from quantum resource-estimation tools can be used as inputs to the runtime model introduced in
  Section~\ref{sec:metrics}.

\begin{table}[!tp]
\tablebodyfont
\caption{Mapping resource-estimator outputs into the runtime model.}
\label{tab:resource-estimator-map}
\centering
\begin{tabularx}{\linewidth}{@{}>{\raggedright\arraybackslash}X l@{}}
\toprule
\textbf{Estimator output} & \textbf{Model input} \\
\midrule
Logical depth or gate count & $C_Q$ \\
Logical cycle time or throughput & $\tau_Q$ \\
Estimated quantum runtime & $T_Q$ \\
Shot count and batching assumptions & $C_Q$, $F$ \\
Measurement or output payload & $V$ \\
Classical post-processing estimate & $C_C$ or $T_C$ \\
\bottomrule
\end{tabularx}
\end{table}

To enable comparison across workflows, we map each metric to a discrete 1--5 scale anchored to representative numerical ranges (Table~\ref{tab:metric-scale}). These ranges are intended as order-of-magnitude guidelines; absolute values depend on the specific problem and hardware.

\begin{table*}[!tp]
\tablebodyfont
\centering
\caption{1--5 scale for hybrid quantum--classical workflow metrics, with representative numerical ranges.}
\label{tab:metric-scale}
\renewcommand{\arraystretch}{1.3}
\begin{tabularx}{\linewidth}{>{\raggedright\arraybackslash}p{2.4cm} c >{\raggedright\arraybackslash}p{3.2cm} >{\raggedright\arraybackslash}X}
\toprule
\textbf{Metric} & \textbf{Score} & \textbf{Range} & \textbf{Description} \\
\midrule

\multirow{5}{*}{\parbox{2.4cm}{Quantum compute cost \\ ($C_Q \propto d \cdot s$)}}
& 1 & $d \sim 10^1$, $s \sim 10^1$ & Very shallow circuits, few shots; toy-scale \\
& 2 & $d \sim 10^2$, $s \sim 10^3$ & Shallow depth, limited shots \\
& 3 & $d \sim 10^2\text{--}10^3$, $s \sim 10^6$ & Moderate depth and shots; representative NISQ workloads \\
& 4 & $d \sim 10^3\text{--}10^4$, $s \sim 10^6\text{--}10^8$ & Deep circuits and/or large shot counts \\
& 5 & $d > 10^4$, $s > 10^8$ & Fault-tolerant scale or extremely shot-intensive \\
\midrule

\multirow{5}{*}{\parbox{2.4cm}{Classical compute cost \\ ($C_C$)}}
& 1 & $T_C < 10^0$ s & Minimal processing; laptop-scale \\
& 2 & $T_C \sim 10^0\text{--}10^1$ s & Single workstation or small server \\
& 3 & $T_C \sim 10^1\text{--}10^2$ s & GPU-accelerated server \\
& 4 & $T_C \sim 10^2\text{--}10^3$ s & Multi-node HPC \\
& 5 & $T_C > 10^3$ s & Distributed HPC or cloud-scale \\
\midrule

\multirow{5}{*}{\parbox{2.4cm}{Communication frequency \\ ($F$ per cycle)}}
& 1 & $F = 1$ & Single exchange per cycle \\
& 2 & $F \sim 10^0\text{--}10^1$ & Few exchanges per cycle \\
& 3 & $F \sim 10^1\text{--}10^2$ & Regular exchanges; hybrid loop \\
& 4 & $F \sim 10^2\text{--}10^3$ & Many exchanges; strong interdependence \\
& 5 & $F > 10^3$ & Near-continuous interaction \\
\midrule

\multirow{5}{*}{\parbox{2.4cm}{Data volume per exchange \\ ($V$)}}
& 1 & $V < 10^1$ B & Bytes to tens of bytes \\
& 2 & $V \sim 10^3$ B & Kilobytes per exchange \\
& 3 & $V \sim 10^5\text{--}10^6$ B & Hundreds of kB to a few MB \\
& 4 & $V \sim 10^7$ B & Tens of megabytes \\
& 5 & $V > 10^8$ B & Hundreds of megabytes or more \\
\midrule

\multirow{5}{*}{\parbox{2.4cm}{Communication-to-computation ratio \\ ($R_{cc}$)}}
& 1 & $R_{cc} < 0.01$ & Compute-dominated; latency negligible \\
& 2 & $R_{cc} \sim 0.01\text{--}0.1$ & Latency has minor impact ($<$10\% runtime) \\
& 3 & $R_{cc} \sim 0.1\text{--}1$ & Latency and compute comparable \\
& 4 & $R_{cc} \sim 1\text{--}10$ & Latency significantly affects performance \\
& 5 & $R_{cc} > 10$ & Communication-dominated; runtime set by $L$ and $V/B$  \\

\bottomrule
\end{tabularx}
\end{table*}

Applying this scale to the workflows analyzed in the main text yields the profiles in Table~\ref{tab:metric-profiles}. All scores are per compute cycle.

\begin{table}[!tp]
\tablebodyfont
\centering
\caption{Metric profiles (1--5 scale) for the workflows in Section \ref{sec:examples} and Appendix~\ref{sec:appendix-sqsp}. $R_{cc}$ for SQSP-MaF reflects the application-level analysis.}
\label{tab:metric-profiles}
\begin{tabular}{lccccc}
\toprule
\textbf{Workflow} & $C_C$ & $C_Q$ & $F$ & $V$ & $R_{cc}$ \\
\midrule
SQD       & 5 & 3 & 1 & 5 & 1 \\
GQE       & 2 & 3 & 1 & 3 & 1 \\
DMRG--QPE & 2 & 5 & 1 & 5 & 1 \\
QE-MCMC   & 1 & 1 & 1 & 1 & 5 \\
SQSP-MaF  & 1 & 2 & 3 & 1 & 4 \\
\bottomrule
\end{tabular}
\end{table}

\begin{figure}[!tp]
    \centering
    \includegraphics[width=\columnwidth]{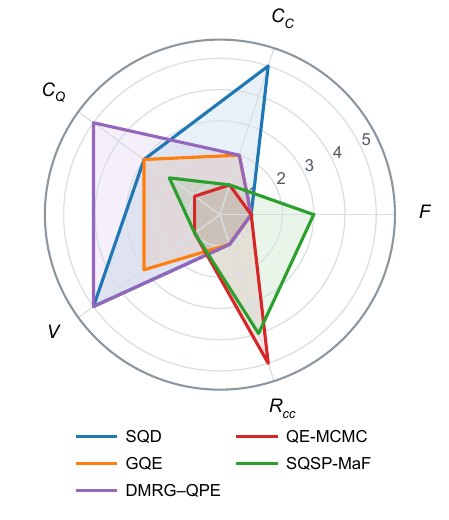}
    \caption{The radar chart compares classical and quantum compute costs ($C_C$, $C_Q$), communication frequency ($F$), data volume ($V$), and communication-to-computation ratio ($R_{cc}$) for sample-based quantum diagonalization (SQD), Generative Quantum Eigensolver (GQE), density matrix renormalization group-assisted quantum phase estimation (DMRG--QPE), sparse quantum state preparation with mid-circuit measurement and feed-forward (SQSP-MaF), and quantum-enhanced Markov chain Monte Carlo (QE-MCMC). Larger radii indicate higher scores on the 1--5 scale in Table~\ref{tab:metric-profiles}; higher $R_{cc}$ indicates greater communication overhead relative to computation.}
    \label{fig:spider}
\end{figure}

Figure~\ref{fig:spider} provides a visual summary of the workflow profiles in Table~\ref{tab:metric-profiles}. It is intended as a
  comparative guide rather than a quantitative runtime estimate: larger radial extent indicates a larger value of the corresponding metric.

\begin{figure}[!tp]
      \centering
      \includegraphics[width=\linewidth]{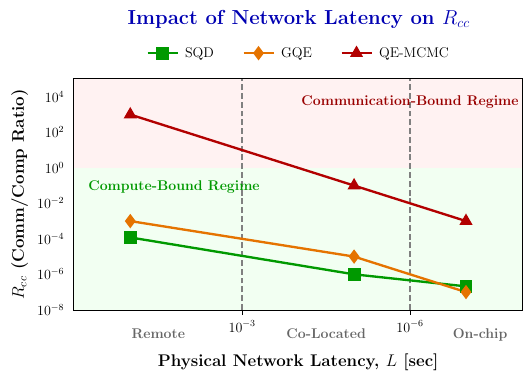}
      \caption{Impact of physical network latency $L$ on the communication-to-computation
      ratio $R_{cc}$ for SQD, GQE, and QE-MCMC across the remote, co-located, and co-designed
      integration tiers. The shaded regions mark the communication-bound ($R_{cc} > 1$)
      and compute-bound ($R_{cc} < 1$) regimes. SQD and GQE remain compute-bound at every
      tier, as their large per-cycle compute costs suppress communication overhead even at
      remote-level latencies. In contrast, QE-MCMC is communication-bound under remote
      access ($R_{cc} \sim 10^3$) and crosses into the compute-bound regime in the co-located and co-designed tiers, consistent with the per-tier estimates in
      Section~\ref{sec:examples}.}
      \label{fig:rcc-network}
  \end{figure}

Figure~\ref{fig:rcc-network} examines the communication-to-computation ratio $R_{cc}$ explicitly, showing how it varies with the physical network latency $L$ between the quantum and classical processors. Whereas the profiles above score $R_{cc}$ at a fixed integration setting, sweeping $L$ across the remote, co-located, and co-designed tiers reveals which workflows change regime as $L$ decreases. 

The hardware-dependent parameters $\tau_Q$, $L$, and $B$ that enter the runtime model vary substantially across QPU modalities. Table~\ref{tab:hardware_metrics} provides representative values for current hardware. Here $\tau_Q$ is the effective quantum throughput (encompassing gate speed, measurement, reset, and control overhead), $F_{\mathrm{gate}}$ is the representative gate fidelity, $L$ is the characteristic round-trip latency of the control system, and $B$ is the sustained interconnect bandwidth.

\begin{table*}[!tp]
\tablebodyfont
\centering
\caption{Representative hardware-dependent parameters for different QPU modalities. Photonic values are largely architecture-dependent and not yet broadly characterized$^\dagger$.}
\label{tab:hardware_metrics}
\begin{tabular}{lcccc}
\toprule
\textbf{Modality} &
$\tau_Q$ &
$F_{\mathrm{gate}}$ &
$L$ &
$B$ \\
\midrule
Superconducting\cite{lacroix2025scaling, ali2024reducing} &
$10^{5}$--$10^{7}$ layers/s &
0.99--0.999 &
$10^{2}$ ns &
GB/s \\
Ion trap\cite{ransford2025helios, myerson2008high-fidelity} &
$10^{3}$--$10^{4}$ layers/s &
0.995--0.9995 &
$10^{2}$ $\mu$s &
MB/s \\
Neutral atom\cite{saffman2025quantum} &
$10^{4}$--$10^{6}$ layers/s &
0.96--0.995 &
$1$--$10$ $\mu$s &
MB/s \\
Photonic$^\dagger$ &
--- &
$0.991$--$0.993$ &
--- &
--- \\
\bottomrule
\end{tabular}

\smallskip
{\footnotesize $^\dagger$Photonic architectures use measurement-based quantum computation (MBQC) rather than a gate-based model; throughput, latency, and bandwidth characteristics depend strongly on the specific architecture and are not yet broadly benchmarked.}
\end{table*}

\section{Decoder Latency and Error Rates}
\label{sec:appendix-decoding}

Fault-tolerant quantum computation is more robust to slow decoding times when compared to non-fault tolerant operation. However, in the extreme limit, even syndrome extraction cannot extend a logical qubit's lifespan forever, and increasing the lifespan via increased code distance is a costly use of resources. Additional trade-offs can be made, such as post-corrected T-gates~\cite{LitinskiGameOfSurfaceCodes} which shift the corrective action to logical ancilla qubits. Ref.~\cite{RiverlaneParallelTime} attempts to alleviate the reduction in logical clock speed by using autocorrected $\pi/8$ PPMs (see Ref.~\cite{LitinskiGameOfSurfaceCodes}, Figs.7, 17b, 25b). However, their analysis is only correct for the case of commuting $\pi/8$ PPMs. For non-commuting operations, the autocorrection qubits must be measured at intervals of $\tau_r$, which means that autocorrected qubits are accumulated much more quickly than they can be measured out. 

While FTQC can be robust to decoding latency in some cases, it is still a key figure of merit, which has dramatic impact on the performance of a quantum computer~\cite{khalid2025impacts}. Decoding must be able to keep up with the syndrome data generation rate at minimum. Lower-latency decoding means that branch decisions on which logical gate to apply can be decided faster. 

As a simple model, we can view the impact of the decoder latency as an idling penalty. Even as we perform repeated rounds of syndrome extraction to preserve logical qubit information, we can quantify this penalty of logical idling as the logical error rate per syndrome extraction round. This sets a characteristic timescale, unique to each device, for the cost of idling. When the physical error rate is below threshold, high code distances can prolong logical qubit lifetimes well beyond physical qubit decoherence timescales. 

To model this idle error, we take the logical error rate per cycle $\varepsilon_L$, and apply it each cycle $r$.
\begin{align}
    P_L(r) &= \frac{1}{2}[1-(1-2\varepsilon_L)^r],
    \label{eq:logical_error_prob}
\end{align}

For small $\varepsilon_L$, this is approximately linear: $P_L \approx \varepsilon_L \cdot r$. While the logical error rate can be measured experimentally, it improves once the physical error rate $p$ falls below the threshold $p_{th}$~\cite{Fowler_2012}:
\begin{align}
    \varepsilon_L &= A \left( \frac{p}{p_{\rm th}} \right)^{\lfloor (d+1)/2 \rfloor}
    \label{eq:LER_per_cycle}
\end{align}

\begin{figure}[!tp]
    \centering
    \includegraphics[width=\columnwidth]{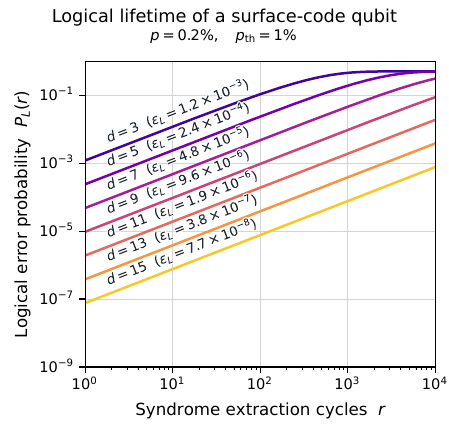} 
    \caption{Logical error probability $P_L(r)$ (Eq.~\ref{eq:logical_error_prob}) as a function of the number of syndrome-extraction cycles $r$, shown for several code distances $d$ with $\varepsilon_L$ given by Eq.~\ref{eq:LER_per_cycle} in the sub-threshold regime ($p < p_{\mathrm{th}}$). Higher code distances suppress the per-cycle logical error rate, extending the usable idle window — the regime in which decoder latency can be absorbed without logical failure.}
    \label{fig:logical_error_rate_vs_cycles}
\end{figure}

Figure~\ref{fig:logical_error_rate_vs_cycles} shows $P_L(r)$ from Eq.~\ref{eq:logical_error_prob} as a function of cycle count $r$ for several code distances, with $\varepsilon_L$ set by Eq.~\ref{eq:LER_per_cycle}: increasing $d$ flattens the curve, extending the number of cycles a logical qubit can idle before logical failure becomes likely.

It is important to note that the trade space is large for achieving fault-tolerant quantum computation. In the ``algorithmic fault-tolerance"~\cite{zhou2025algorithmic}, transversal operations and correlated decoding lead to a reduction in the number of syndrome extraction cycles needed. 

\section{Quantum Memory}
\label{sec:appendix-qmem}

A useful general-purpose quantum computing system may very well require a quantum memory and will naturally converge toward a quantum version of the Von Neumann architecture. In the classical Von Neumann model, instructions and data reside in a unified memory space and are accessed over a shared bus by the processing unit. This stored program concept enables general-purpose computation and flexible reprogramming by modifying memory contents, at the cost of a well-known bandwidth and latency bottleneck between compute and memory. Despite this limitation, this model is the foundation for modern classical computing, with tightly coupled components such as the CPU or GPU, main memory, input output, and a shared interconnect. Here, quantum memory means coherent storage that preserves quantum states across stages of an algorithm. This is distinct from QRAM, which is a more specialized model for coherently accessing stored data by address.

In a quantum Von Neumann–style architecture, quantum memory functions as a coherent extension of the QPU, enabling quantum states to persist across distinct phases of an algorithm. The key feature is the QPU–quantum-memory coupling, which allows intermediate quantum states to be stored and later reintroduced into quantum computation without re-preparation. This capability shifts quantum execution away from a strictly stateless circuit model toward stateful quantum workflows, where quantum data survives beyond a single kernel call.

Within this execution model, low-latency classical–quantum interconnects enable efficient orchestration of stateful quantum workflows. While quantum data remains resident in quantum memory, classical accelerators can perform nontrivial computation to determine subsequent quantum operations like conditional calculations. The interconnect therefore supports rapid resumption and transformation of stored quantum states, rather than repeated state initialization, which is a bottleneck of many NISQ algorithms.

The availability of quantum memory could enable new algorithmic structures that are inaccessible in today's execution model. These include multi-stage quantum programs with intermediate state checkpointing, conditional execution paths acting on preserved quantum states, improved preservation of expensive to prepare quantum states across algorithmic phases, and separation of quantum coherence time from classical decision latency. More broadly, quantum memory enables a stored-state programming model for quantum algorithms, analogous in spirit to the stored-program concept in classical computing, and represents a key element for the future architecture of quantum computers. However, a quantum memory still would suffer from finite lifetimes, though can be designed to target improved logical lifespans, most likely at the cost of worsened computational capability.

Within our framework, quantum memory has direct implications for several metrics. By allowing quantum states to persist across algorithmic phases, it can restructure the compute cycle itself - enabling workflows where expensive state preparation is amortized over multiple uses rather than repeated each cycle, effectively reducing $C_Q$ per cycle. It can also reduce the communication frequency $F$ by eliminating round trips that would otherwise be needed to re-initialize quantum states from classical descriptions. Perhaps most significantly, quantum memory relaxes the requirement that classical decision latency fit within the coherence time of the computational qubits: the classical host can perform arbitrarily complex computation ($T_C$) while the quantum state remains stored, decoupling $R_{cc}$ from the physical coherence window. This would shift workflows that are currently real-time level into the application-level regime.

\section{Workflow Families}
\label{sec:appendix-families}
Many quantum algorithms share core primitives and control-flow patterns, leading to similar metric profiles. Table~\ref{tab:workflow_class} proposes a grouping of some common workflows into families that broadly share characteristics, along with a general expectation on the communication-to-computation regime. Any grouping of quantum algorithms is unlikely to cover all use cases, but this illustrates that often a representative member of a family can serve as a benchmark for the integration requirements of other members. 

\begin{table*}[!tp]
\tablebodyfont
\centering
\caption{Quantum algorithms grouped into broad families that generally share resource characteristics, with representative examples for each family. These groupings are intended to highlight similarities among methods and guide thinking about quantum algorithms, though not all algorithms fit neatly into the proposed categories.}
\label{tab:workflow_class}
\renewcommand{\arraystretch}{1.15}
\setlength{\tabcolsep}{6pt}
\begin{tabularx}{\linewidth}{@{}
    >{\hsize=0.9\hsize\linewidth=\hsize\raggedright\arraybackslash}X
    >{\hsize=1.6\hsize\linewidth=\hsize\raggedright\arraybackslash}X
    >{\hsize=0.5\hsize\linewidth=\hsize\raggedright\arraybackslash}X@{}}
\toprule
\textbf{Algorithm family} & \textbf{Examples} & \textbf{$R_{cc}$ regime} \\
\midrule
Variational quantum circuits
 & VQE, ADAPT-VQE \& variants\newline
   GQE \& variants~\cite{GPT-QE, tyagin2025qaoagpt}\newline
   Quantum Kernels
 & Low \\
\midrule
Quantum-enhanced classical algorithms
 & SQD, SQKD\newline
   QC-AFQMC
 & Very low \\
\midrule
Matrix function computation
 & QPE\newline
   HHL\newline
   QSVT
 & Low \\
\midrule
Time evolution
 & Trotterization\newline
   Stochastic (qDRIFT)\newline
   (Pseudo) Adiabatic
 & Very low \\
\midrule
Non-variational optimization
 & Decoded Quantum Interferometry~\cite{jordan2025optimization}\newline
   Quantum Semi-Definite Programming~\cite{brandao2017quantum}
 & Very low \\
\midrule
Feed-forward algorithms
 & QCNN~\cite{cong2019quantum}\newline
   Adaptive Circuit Knitting~\cite{adaptiveknitting26}\newline
   IQPE~\cite{dobsiceketal2007}\newline
   Quantum Metropolis Sampling~\cite{jiang2024quantum}\newline
   Local Alternating Quantum Classical Computations~\cite{Buhrman2024state}
 & Moderate to high \\
\bottomrule
\end{tabularx}
\end{table*}

As a quantitative example for the matrix function computation family, Table~\ref{tab:hubbard_rcc} gives illustrative $R_{cc}$ values for ground-state energy estimation of the 2D Hubbard model at three lattice sizes. It targets an energy error of $10^{-2}t$ per site and assumes a remote integration tier ($L=100$~ms, $B=1$~Gbps), $10^5$ T gates per second, and 4 T gates per Toffoli. For DMRG state preparation, we fit the bond dimension to the $SU(2)$ DMRG data of~\cite{liu2025accurate}, giving 170, 1700 and 10000 for the 4x4, 6x6 and 8x8 lattices. We assume the MPS can be converted to $U(1)$ form and truncated back to this bond dimension. The Toffoli cost of preparing it follows Appendix B of~\cite{rupprecht2026faster} with $b=15$. The classical cost is $N_{iter}N_{sys}\chi^3$ operations ($N_{iter}=100$) at $10^{12}$ FLOP/s, and $N_{sys}\chi^2$ parameters are sent to the QPU. For qubitization-based QPE, we take the leading-order T count from~\cite{kan2025resource-optimized}, budget 10~MB for the circuit specification, neglect classical cost, and assume a suitable initial state. Both estimates neglect mid-circuit measurements.

In all cases $R_{cc}$ is at most a few percent, so even a remote tier adds little to the runtime. For QPE, this is consistent with the low $R_{cc}$ regime assigned to matrix function computation in Table~\ref{tab:workflow_class}, and $R_{cc}$ falls further as the lattice grows, because the circuit specification stays fixed while the QPE cost grows. For DMRG state preparation it stays at a few $\times10^{-3}$, because the transferred MPS grows with the bond dimension along with the compute cost.

\begin{table}[h]
\centering
\caption{Example $R_{cc}$ values for various sizes of the Hubbard model with different algorithms, assuming remote integration tier. See text for estimation details, data estimated from the literature for illustrative purposes. DMRG is the cost of DMRG state preparation, qubitization that of QPE with qubitization.}
\label{tab:hubbard_rcc}
\small
\newcolumntype{L}{>{\raggedright\arraybackslash}X}
\begin{tabularx}{\columnwidth}{@{}lLLL@{}}
\toprule
\textbf{Lattice size} & 4x4 & 6x6 & 8x8 \\
\midrule
DMRG & $3.5\cdot10^{-3}$ & $3.0\cdot10^{-3}$ & $6.4\cdot10^{-3}$ \\
qubitization & $4.6\cdot10^{-2}$  & $2.1\cdot10^{-2}$  & $1.2\cdot10^{-2}$  \\
\bottomrule
\end{tabularx}
\end{table}

\section{Additional Workflows}
\label{sec:appendix-workflows}

\paragraph{Dynamic Stateprep: SQSP-MaF}
\label{sec:appendix-sqsp}
SQSP-MaF is a dynamic-circuit variant of Sparse Quantum State Preparation (SQSP) that uses mid-circuit measurement and classical feed-forward to prepare a target $n$-qubit state with $k$ nonzero amplitudes~\cite{lu2025dynamic}. A static implementation requires circuit depth $O(n \log k)$; the MaF variant reduces this to $O(n)$ by replacing high-depth operations (fan-out gates, OR-controlled X gates) with constant-depth MaF circuits, each consisting of one round of mid-circuit measurement, classical computation on the outcomes, and conditional single-qubit corrections. The classical computation consists of boolean operations on measurement outcomes and executes in nanoseconds on classical control hardware. The $n$ target qubits are processed sequentially, with each step requiring one such MaF round. A compute cycle is one complete state preparation, so there are F = $n$ communication events per compute cycle. Each communication event only transmits $O(k)$ single-bit outcomes and simple gate instructions. As a result of the low classical and quantum compute costs, $R_{cc}$ is moderate to high at the application level. 

\begin{table}[!tp]
\tablebodyfont
\centering
\caption{Metric profile for SQSP-MaF, reported per compute cycle (one state preparation, comprising $n$ sequential MaF rounds); the application-level $R_{cc}$ row is shown for comparison. Representative values assume a transmon platform, with the MaF-round time estimate drawn from~\cite{CarreraVazquez2024dynamic}.}
\label{tab:sqsp-metrics}
\small
\newcolumntype{L}{>{\raggedright\arraybackslash}X}
\begin{tabularx}{\linewidth}{@{}lL@{}}
\toprule
\textbf{Metric} & \textbf{Value} \\
\midrule
$C_C$ & Very low \newline
${\sim}$ns on classical control hardware \\
$C_Q$ & $O(n)$ gate depth, single shot \newline
${\sim}n \times 0.5\,\mu$s~\cite{CarreraVazquez2024dynamic} cycle runtime \\
$F$ & $n$ \\
$V$ & $O(k)$ for measurements and gate instructions \newline
${\sim}1$--$10$~B \\
$R_{cc}$ & Moderate to high at application level \newline
Limited by $\tau_{\mathrm{phys}}$ at real-time level\\
\midrule
\multicolumn{1}{@{}l}{Bottleneck} & Qubit coherence window \\
\bottomrule
\end{tabularx}
\end{table}

However, the application-level diagnostic does not capture the binding constraint in the NISQ regime. Between each measurement and the subsequent feed-forward operation, the unmeasured qubits idle and decohere. The relevant condition is not whether communication overhead is a large fraction of runtime, but whether the classical round-trip fits within the qubit coherence window. This is the feasibility condition of Eq.~\ref{eq:feasibility}, extended by $T_{Q,\mathrm{step}}$ since the quantum operations of each MaF round also consume the coherence window:
\begin{equation}
    T_{Q, \mathrm{step}} + T_{C,\mathrm{step}} + T_{\mathrm{comm,step}} < \tau_{\mathrm{phys}},
\end{equation}
where $\tau_{\mathrm{phys}}$ is the effective idle lifetime of the physical qubits.

This is where the real-time level analysis diverges most sharply from the application level: the specific hardware modality can no longer be abstracted away. For transmons ($\tau_{\mathrm{phys}} \sim 10$--$100\;\mu$s), demonstrated feed-forward latencies of ${\sim}0.5\;\mu$s~\cite{CarreraVazquez2024dynamic} leave workable but narrow margin; trapped ions relax this deadline by orders of magnitude ($\tau_{\mathrm{phys}} \sim 1$--$10$~s) at the cost of slower gates; neutral-atom mid-circuit measurement and feed-forward remain experimentally open. Whether the feasibility condition is comfortably satisfied, marginal, or unattainable thus depends on the modality-specific interplay of coherence times, gate speeds, and feed-forward latency. Our real-time level metrics for SQSP-MaF are summarized in Table \ref{tab:sqsp-metrics}.

On fault-tolerant hardware, where logical qubits are stabilized by the error-correction substrate, the decoherence deadline is relaxed and the workflow reverts to a standard application-level analysis.

\paragraph{Iterative Quantum Phase Estimation (IQPE)}
In IQPE~\cite{dobsiceketal2007}, a compute cycle corresponds to extracting the full $m$-bit phase estimate via $m$ sequential rounds of controlled-unitary application, ancilla measurement, and classical feedback.

\begin{itemize}
    \item \textbf{Classical Compute Cost ($C_C$):} Very low. Simple arithmetic to compute each feedback rotation angle.

    \item \textbf{Quantum Compute Cost ($C_Q$):}
    \begin{itemize}
        \item \textbf{Gate Depth:} High. Each precision bit doubles the controlled-unitary applications, scaling as $O(1/\varepsilon)$. Gate depths can reach ${\sim}10^6$ for chemistry.
        \item \textbf{Sample Complexity:} One measurement per round; overall repetitions scale as $O(1/p^2)$ for overlap $p$ with the target state.
    \end{itemize}

    \item \textbf{Communication frequency ($F$):} $F = m$ per cycle, one blocking exchange per precision bit.

    \item \textbf{Data Volume per Exchange ($V$):} Very low --- a single bit plus a rotation angle per exchange.

    \item \textbf{Communication-to-computation ratio ($R_{cc}$):} Moderate. Between each exchange, at least one controlled-unitary application provides a computational buffer that reduces the effective communication-to-computation ratio relative to workflows like QE-MCMC.
\end{itemize}

  \end{appendices}
\end{document}